\documentclass[showpacs,twocolumn,floatfix,amsmath,amssymb,superscriptaddress,prl,nopacs]{revtex4-1}
\usepackage[utf8]{inputenc}
\usepackage{epsfig}
\usepackage{graphicx}
\usepackage{longtable}
\usepackage{color}
\usepackage{dcolumn}
\usepackage{bm}
\usepackage[colorlinks=true,citecolor=magenta,linkcolor=blue]{hyperref}
\usepackage{dsfont}

\newcommand{\bea}{\begin{eqnarray}}
\newcommand{\eea}{\end{eqnarray}}
\newcommand{\bt}{\textbf}

\newcommand{\phd}{\phantom{\dag}}
\newcommand{\ph}{\phantom{.}}

\newcommand{\noi}{\noindent}
\newcommand{\no}{\nonumber}

\begin{document}

\title{Synthetic Weyl Points and Chiral Anomaly in Majorana Devices\\ with Nonstandard Andreev-Bound-State Spectra}
\author{Panagiotis Kotetes}
\email{kotetes@itp.ac.cn}
\affiliation{CAS Key Laboratory of Theoretical Physics, Institute of Theoretical Physics, Chinese Academy of Sciences, Beijing 100190, China}
\affiliation{Niels Bohr Institute, University of Copenhagen, 2100 Copenhagen, Denmark}
\author{Maria Teresa Mercaldo}
\affiliation{Dipartimento di Fisica ``E. R. Caianiello", Universit\`a di Salerno, IT-84084 Fisciano (SA), Italy}
\author{Mario Cuoco}
\affiliation{CNR-SPIN, IT-84084 Fisciano (SA), Italy}
\affiliation{Dipartimento di Fisica ``E. R. Caianiello", Universit\`a di Salerno, IT-84084 Fisciano (SA), Italy}

\begin{abstract}
We demonstrate how to design various nonstandard types of Andreev-bound-state (ABS) dispersions, via a composite construction relying on Majorana bound states (MBSs). Here, the MBSs appear at the interface of a Josephson junction consisting of two topological superconductors (TSCs). Each TSC harbors multiple MBSs per edge by virtue of a chiral or unitary symmetry. We find that, while the ABS dispersions are $2\pi$-periodic, they still contain multiple crossings which are protected by the conservation of fermion parity. A single junction with four interface MBSs and all MBS couplings fully controllable, or, networks of such coupled junctions with partial coupling tunability, open the door for topological bandstructures with Weyl points or nodes in synthetic dimensions, which in turn allow for fermion-parity (FP) pumping with a cycle set by the ABS-dispersion details. In fact, in the case of nodes, the FP pumping is a manifestation of chiral anomaly in 2D synthetic spacetime. The possible experimental demonstration of ABS engineering in these devices, further promises to unveil new paths for the detection of MBSs and higher-dimensional chiral anomaly. 
\end{abstract}

\maketitle

Superconducting devices are currently in the spotlight~\cite{Castelvecchi}, with Josephson junctions of conventional superconductors (SCs) being core for state-of-the-art quantum-computing circuits~\cite{Makhlin,Devoret}. Each junction is characterized by the super\-con\-duc\-ting phase dif\-fe\-ren\-ce $\Delta\phi$ across it, and the as\-so\-cia\-ted Josephson ener\-gy scale $E_J(\Delta\phi)$~\cite{Josephson,AB}. When the Josephson link is a quantum-point contact or a quantum dot, the resulting transport~\cite{KO,Furusaki,Carlo,FurusakiRev} is mainly me\-dia\-ted by a small number of electronic so-called Andreev bound states (ABSs)~\cite{Andreev}, which are loca\-li\-zed near the interface. For a junc\-tion of two conventional SCs, the ABS energy dispersions come in pairs~\cite{Furusaki,Carlo,FurusakiRev} $\varepsilon(\Delta\phi)=\pm E_{\rm ABS}\sqrt{1-{\cal T}\sin^2(\Delta\phi/2)}$, with ${\cal T}$ de\-no\-ting the tran\-sparency of the junction and $E_{\rm ABS}$ an ener\-gy scale usual\-ly coinciding with the bulk gap of the SCs~\cite{Zazunov}. For a perfectly transparent junction, the ABS dispersions exhibit a li\-near cros\-sing at $\Delta\phi=\pi$~\cite{KO}. This is not protected, and a gap opens for ${\cal T}\neq1$. In the high-transparency and low-inductance regime of the circuit, a pair of ABS levels can be employed to define a qubit~\cite{Zazunov}, with $\Delta\phi=\pi$ being the operational sweet-spot value. 

Perspectives for ABS engi\-nee\-ring and novel quantum-computing architectures open up in junctions of topological superconductors (TSCs). These may be either intrinsic, e.g., p-wave SCs~\cite{Sigrist,TanakaPWave,ReadGreen,Ivanov,Salts,KitaevUnpaired,Sengupta,VolovikBook,Maeno2012,purple,Dumi1,Wakatsuki,Dumi2,Mercaldo2016,Mercaldo2017,SatoAndo}, or artificial~\cite{FuKane,HasanKane,QiZhang,Alicea,CarloRev,Leijnse,KotetesClassi,Franz,Aguado,LutchynNatRevMat}, arising in hybrid devices~\cite{SauPRL,AliceaPRB,LutchynPRL,OregPRL,Choy,KarstenNoSOC,NadgPerge,Nakosai,Braunecker,Klinovaja,Vazifeh,Pientka,Ojanen,Heimes,Brydon,Li,Heimes2,JinAn,Mike,Silas,Andolina}. Both types harbor zero-energy and charge-neutral Majorana bound states (MBSs)~\cite{ReadGreen,Ivanov,KitaevUnpaired}. Remar\-ka\-bly, a Josephson junc\-tion of two TSCs features pairs of coupled MBSs which give rise to $4\pi$-periodic ABS ener\-gy dispersions~\cite{KitaevUnpaired,Yakovenko,FuKaneJ,TanakaYokohamaNagaosa,CarloJ,KotetesJ,AliceaJ,Sticlet2013,PientkaJ,Cayao}, i.e., $\propto\cos(\Delta\phi/2)$. In contrast to the conventional s-wave and unconventional d-wave~\cite{TanakaDWave,TanakaReview} ABS cases, here, the li\-near crossing at $\Delta\phi=\pi$ is protected by the conservation of fermion pa\-ri\-ty (FP)~\cite{FuKaneJ,CarloJ}, i.e., the number of electrons in the system modulo two. The exotic properties of the MBSs, and their promise for fault-tolerant quantum com\-pu\-ting~\cite{Ivanov,KitaevTQC,Nayak,AliceaTQC,JianLi}, sparked recently their intense pursuit and resulted in their spectroscopic detection in va\-rious artificial platforms~\cite{Mourik,Yazdani1,Yacoby,ShupingLee,Ruby,Meyer,Jinfeng,Molenkamp,MT,Sven,Fabrizio,SCZhang,Yazdani2,Giazotto,HaoZhang,Attila,Wiesendanger,Tonio,Ren}. The next milestone is the demonstration of quantum mani\-pu\-la\-tions~\cite{Milestones,Karzig,Tommy} using MBSs. However, this is currently a challen\-ging feat and, hence, it is urgent to unveil new phenomena that take advantage of the topolo\-gi\-cal character of MBSs.

In this Letter, we propose to employ MBSs as buil\-ding blocks for engineering various nonstandard ABS ener\-gy dispersions, which further unlock new types of Berry-phase~\cite{Berry,Niu} and chiral-anomaly effects~\cite{CA}. For this purpose, we consider junctions consisting of two TSCs, with each one har\-boring multiple zero-energy MBSs per edge by virtue of unitary or chiral symmetries~\cite{Altland,KitaevClassi,Ryu,TeoKane,KotetesClassi,Shiozaki,Chiu}. We consider the miminal construc\-tion, where each TSC contributes with two MBSs per edge, i.e. $\gamma_{1,2}$ and $\gamma_{3,4}$. The resulting four MBSs can be pairwise coupled by terms ($t_{12,34}$) vio\-la\-ting the symmetries pro\-tec\-ting the multiple MBSs before the TSCs get in contact, as well as by Majorana-Josephson terms ($t_{13,24,14,23}$) ori\-gi\-na\-ting from electron tunneling across the junction. The latter show a $4\pi$-periodic dependence in terms of the phase difference~\cite{KitaevUnpaired,Yakovenko,FuKaneJ,TanakaYokohamaNagaosa,CarloJ,KotetesJ,AliceaJ,Sticlet2013,PientkaJ,Cayao}, i.e. $t_{nm}\propto\cos\big[(\Delta\phi+\beta_{nm})/2\big]$, with $nm=13,24,14,23$. The phases $\beta_{nm}$ are determined by the symmetries dictating the two TSCs and the interface.

We mainly find $2\pi$-periodic ABS dispersions which, for $t_{14,23}=0$, follow the form $\varepsilon(\Delta\phi)\propto \eta\cos\chi+\cos(\Delta\phi+\varphi)$. This situation takes place when the MBS state vectors differ across the junction only by phase factors set by $\varphi$ and $\chi$, and the interface preserves the symmetries of the two TSCs. The parameter $\eta$ controls the relative strength of the intra- and inter-TSC MBS couplings. Apart from fully-gapped spectra, we also obtain novel types of gapless dispersions, featuring either multiple FP-protected linear crossings or quadratic band touchings.

We show that such ABS spectra can be harnessed to engineer novel types of topologically-nontrivial gapped bandstructures in synthetic spaces (cf. Refs.~\onlinecite{Lewenstein,Nazarov,Levchenko,MeyerHouzet}) containing Berry monopoles~\cite{Berry,Niu}, i.e. points in pa\-ra\-me\-ter space where band touchings occur. These become already accessible in 1D TSC junctions with a full gap. Alternatively, one can employ networks of 1D TSC junctions with gapless spectra to obtain fully-gapped nontrivial bands. The presence of Berry monopoles allows for Cooper pair charge pumping~\cite{Geerligs,Jukka} \`{a} la Thouless~\cite{Thouless}, by simultaneously sweeping $\Delta\phi$ and an additional parameter $\theta$. A full cycle pumps an integer number of $2e$ charge which, however, does not switch the FP. Hence, to enable $1e$-charge transfer and thus FP pum\-ping~\cite{TeoKane,Keselman}, we propose to sweep in $(\Delta\phi,\theta)$ space \textit{selectively}. Thus, the fractional contribution of each Berry monopole can be isolated. Notably, the aspect of selective swee\-ping and its applica\-bi\-li\-ty to TSCs with multiple MBS which are not necessarily Kramers pairs, dif\-fe\-ren\-tiates our $\mathbb{Z}$-FP pump from the $\mathbb{Z}_2$-FP pump of Ref.~\onlinecite{Keselman}.

We now lay out our general approach to the compo\-si\-te construction of the ABSs, by em\-ploying a mi\-ni\-mum of four MBSs which appear near the junction's interface. The correspon\-ding MBS operators satisfy $\{\gamma_n,\gamma_m^{\dag}\}=\delta_{nm}$ with $n,m=1,2,3,4$. Since the MBSs define zero-energy quasiparticle excitations, $\gamma_n=\gamma_n^{\dag}$ for all $n=1,2,3,4$, which further leads to the nonstandard relations $\gamma_n^2=1/2$. Each pair of MBSs appearing on the edge of each TSC is protected by chiral or unitary symmetries. The low-energy Hamiltonian obtained by projecting onto the MBS subspace reads ${\cal H}=\sum_{n,m=1,2,3,4}^{n<m}i\gamma_nt_{nm}\gamma_m$, with the coupling matrix elements $t_{12}$ and $t_{34}$ ($t_{13,24,14,23}$) ha\-ving an intra-TSC (inter-TSC) character. These, ge\-ne\-rally depend on the superconducting phase dif\-fe\-ren\-ce biasing the junction. Here, we restrict to ABS spectra twisted by a charge-phase difference $\Delta\phi$. As we tho\-roughly discuss in our accompanying work in Ref.~\onlinecite{JointMercaldo}, intrinsic spin-triplet p-wave SCs allow for the pos\-si\-bi\-li\-ty of separately imposing a phase dif\-fe\-ren\-ce $\Delta\phi_{\uparrow,\downarrow}$ in each spin sector, which can be split into charge- and spin-phase components.

We re-express the MBS Hamiltonian as follows:
\bea
{\cal H}=\frac{i}{2}\bm{\Gamma}^{\intercal}\hat{B}\bm{\Gamma}=\frac{1}{2}\sum_{s=\pm}\left(\begin{array}{cc}a_s^{\dag}&a_s\end{array}\right)
\left(\begin{array}{cc}\varepsilon_s&0\\0&-\varepsilon_s\end{array}\right)
\left(\begin{array}{cc}a_s\\a_s^{\dag}\end{array}\right)\quad
\label{eq:LowEnergyGeneric}
\eea

\noi where we employed the Majorana multicomponent ope\-ra\-tor $\bm{\Gamma}^{\intercal}=(\gamma_1\,\,\gamma_2\,\,\gamma_4\,\,\gamma_3)$, its transpose $\bm{\Gamma}$, and the skew-symmetric matrix $\hat{B}$. Moreover, we introduced the ABS fermionic ope\-ra\-tors $a_{\pm}$, sa\-ti\-sfying $\{a_s,a_p^{\dag}\}=\delta_{s,p}$ and $\{a_s,a_p\}=0$, with $s,p=\pm$. The ABS dispersions read:
\bea
&&\varepsilon_{+}=\sqrt{{\cal S}+\sqrt{{\cal S}^2-\big[{\rm Pf}(\hat{B})\big]^2}}\phd\ph{\rm and}\ph\phd 
\varepsilon_+\varepsilon_-={\rm Pf}(\hat{B})\,,\ph\quad\label{eq:ABSGenericDispersions}
\eea

\begin{figure}[t!]
\centering
\includegraphics[width=1\columnwidth]{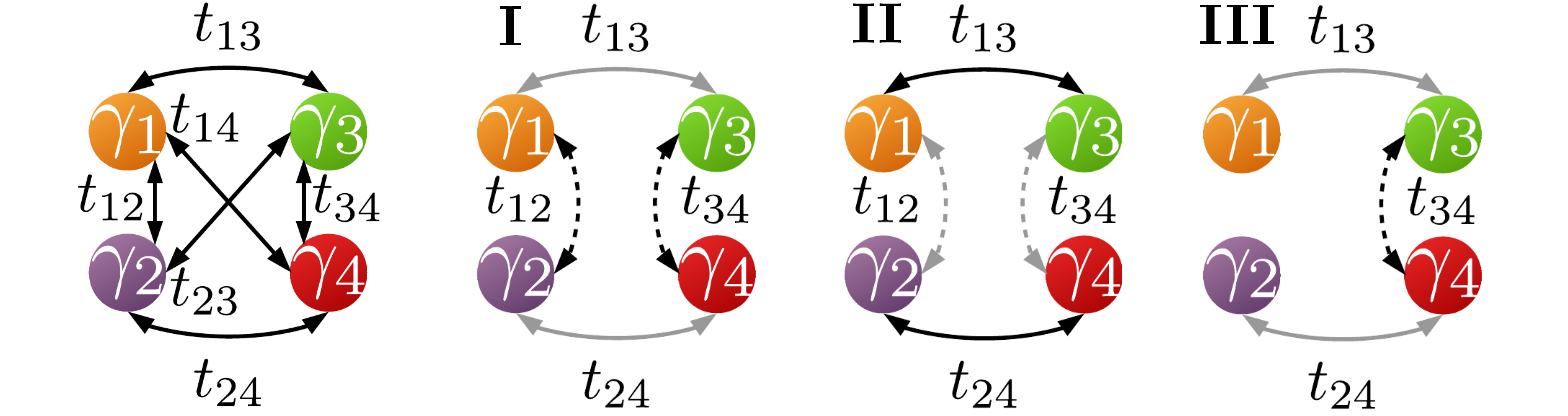}
\protect\caption{Majorana-bound-state (MBS) couplings for four MBSs appearing at the interface of a junction formed by two topological superconductors (TSCs). Each TSC provides a pair of decoupled or hybridized MBSs, before contact. The couplings are divided into intra-TSC $t_{12,34}$ and inter-TSC $t_{13,24,14,23}$. We focus on three representative scenarios I, II and III, depen\-ding on the strengths of the couplings. The dark- and light-color arrows denote the strong and weak MBS couplings. If a line is missing, the respective coupling is zero.}
\label{Fig:Fig1}
\end{figure}

\noi with ${\cal S}=\sum_{n,m=1,2,3,4}^{n<m}t_{nm}^2/2$ and the Pfaffian ${\rm Pf}(\hat{B})=t_{13}t_{24}-t_{12}t_{34}-t_{14}t_{23}$. An equivalent expression was found in Ref.~\onlinecite{Sticlet2013}, which focussed on different aspects of multiple MBSs than the ones considered here. We insist to express $\varepsilon_-$ as in Eq.~\eqref{eq:ABSGenericDispersions}, obtained by $(\varepsilon_+\varepsilon_-)^2=[{\rm Pf}(\hat{B})]^2$, since this form reflects the antisymmetry of the MBS couplings~\cite{KitaevUnpaired}, and further allows for a transparent description of ABS dispersions with protected crossings.

We move on with inferring the ABS dispersions obtained for the three limiting scenarios of Fig.~\ref{Fig:Fig1}, in which, $t_{14,23}=0$, $t_{13}\propto t_{13}^{(0)}\cos\left[\left(\Delta\phi+\varphi+\chi\right)/2\right]$, and 
$t_{24}\propto t_{24}^{(0)}\cos\left[\left(\Delta\phi+\varphi-\chi\right)/2\right]$, with $t_{nm}^{(0)}$ the MBS couplings obtained for $\Delta\phi=0$. Here, the interface preserves the symmetries of the two TSCs before contact, which further implies that the intra-TSC coupling elements $t_{12,34}$ are independent of $\Delta\phi$, i.e. $t_{12}=t_{12}^{(0)}$ and $t_{34}=t_{34}^{(0)}$.

\begin{figure}[t!]
\centering
\includegraphics[width=1\columnwidth]{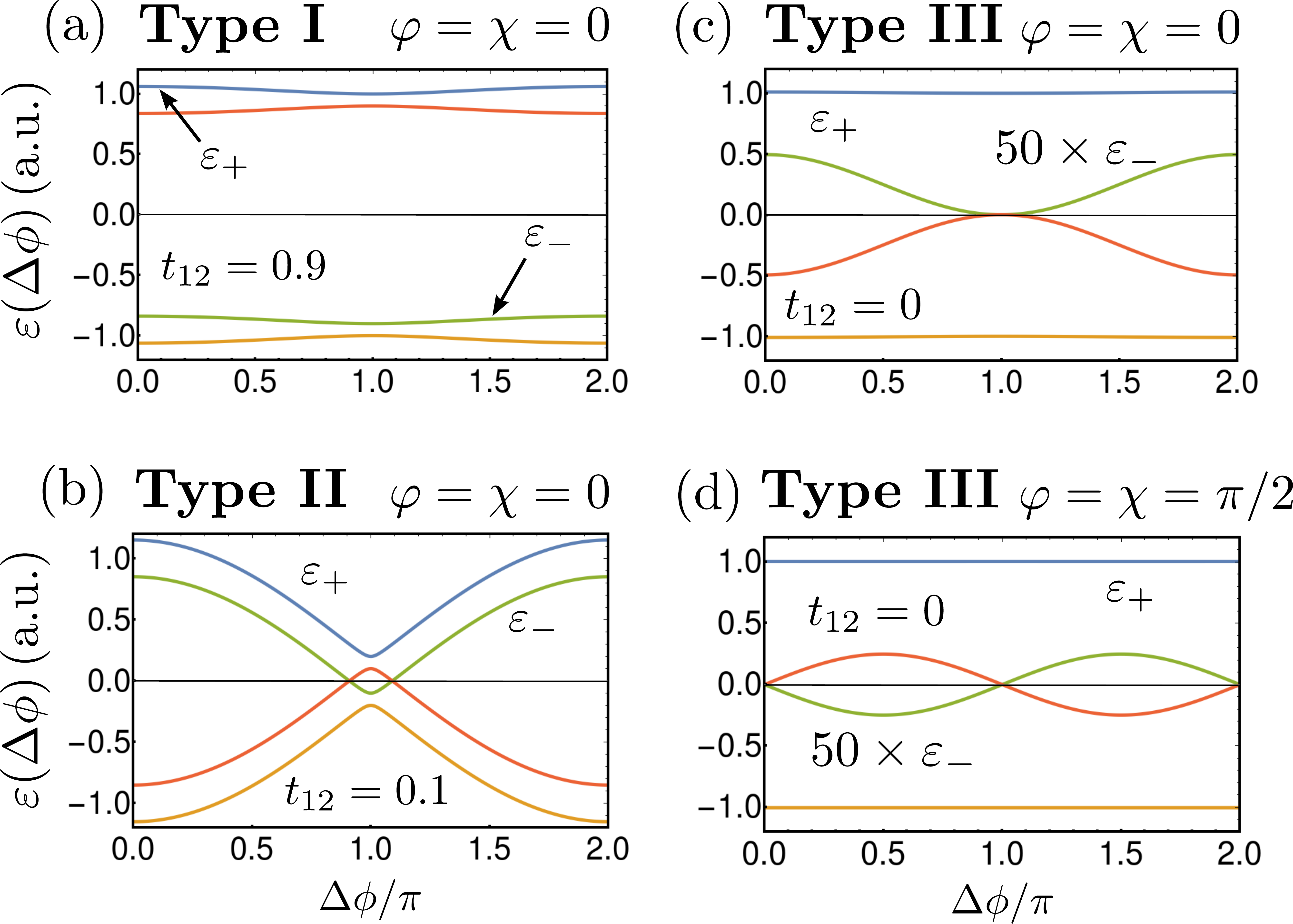}
\protect\caption{ABS spectra corresponding to the three types of scenarios shown in Fig.~\ref{Fig:Fig1}, when the two TSCs are biased by a phase difference $\Delta\phi$. In case (a) we obtain a fully-gapped ABS spectrum, while in case (b), the lowest ABS branch contains two FP-protected linear crossings. Case (c) can be viewed as the critical situation where the two linear crossings of case (b) merge into a single quadratic band crossing. In case (d), the $\varepsilon_-$ branch has a sinusoidal form and contains two linear crossings at $0$ and $\pi$. In (b) $t_{13}^{(0)}=t_{24}^{(0)}=1$ and $t_{34}^{(0)}=0.2$. In the rest $t_{13}^{(0)}=t_{24}^{(0)}=0.1$ and $t_{34}^{(0)}=1$. In all panels $t_{14,23}=0$.}
\label{Fig:Fig2}
\end{figure}

$\bm{|t_{13}^{(0)}|,|t_{24}^{(0)}|\ll|t_{12}^{(0)}|,|t_{34}^{(0)}|}$. In this case, the inter-TSC MBS couplings are considered to be substantially weaker than the intra-TSC ones, a situation which is feasible by re\-du\-cing the junction's transparency. At the same time, $|t_{12}|$ and $|t_{34}|$ are still considered much smaller than the bulk energy gap so that the present low-ener\-gy projection remains valid. Given the above, we find that both pairs of edge MBSs hybridize to ABSs, with ener\-gy dispersions $\varepsilon_{+}\approx t_{34}$ and $\varepsilon_{-}\approx -t_{12}$, for the choice $t_{34}>t_{12}>0$. Since $t_{12}$ and $t_{34}$ are independent of $\Delta\phi$, the ABS bands are very weakly dispersive with $\Delta\phi$. With the use of Eq.~\eqref{eq:ABSGenericDispersions}, we show such a dispersion in Fig.~\ref{Fig:Fig2}(a).

$\bm{|t_{12}^{(0)}|,|t_{34}^{(0)}|\ll|t_{13}^{(0)}|,|t_{24}^{(0)}|}$. We now consider the inverse limit, in which the symmetries protecting the multiple MBSs are weakly violated. In the fully-symmetric $t_{12}=t_{34}=0$ case, each isolated TSC has a symmetry-protected pair of MBSs per edge. After contact, one finds the $4\pi$-periodic~\cite{KitaevUnpaired} ABS dispersions $\varepsilon_+=t_{13}$ and $\varepsilon_-=t_{24}$, with each one containing a single FP-protected cros\-sing~\cite{FuKaneJ,CarloJ}. When $\chi=0$, the two crossings appear for the same value of $\Delta\phi$, and the symmetry-violating $t_{12}$ and $t_{34}$ terms open an energy gap at the dege\-ne\-ra\-cy point. This gap opening redistributes the two initially coin\-ci\-ding cros\-sings, by transferring them to one of the two resulting $2\pi$-periodic ABS dispersions. Fig.~\ref{Fig:Fig2}(b) depicts such a bandstructure, which can be understood by separately examining the ABS dispersions near and away from the degeneracy point. Away from it, the condition $|t_{12}^{(0)}|,|t_{34}^{(0)}|\ll|t_{13}^{(0)}|,|t_{24}^{(0)}|$ implies that $\varepsilon_+\approx t_{13}$ and $\varepsilon_-\approx t_{24}$. Instead, this hie\-rar\-chy becomes inverted near the degeneracy point, where the ABS energies are given by the gap-opening terms $t_{12}$ and $t_{34}$, as in Fig.~\ref{Fig:Fig2}(a). A smooth connection between the two regimes is ensured by the appearance of two linear crossings. These are FP-protected and, unless a gap closing occurs, they are only removable by annihilating each other. This is achievable, for instance, by increasing the intra-TSC MBS couplings.

$\bm{|t_{13}^{(0)}|,|t_{24}^{(0)}|\ll|t_{34}^{(0)}|}$ and $\bm{|t_{12}^{(0)}|=0}$. We proceed by assuming that the two MBSs of the first TSC are uncoupled even after contacting the second TSC, which is assumed to feature a nonzero $t_{34}$, stemming from the vio\-la\-tion of the ensuing symmetry. This violation is either externally imposed or spontaneously chosen by the hybrid system~\cite{JointMercaldo}. Using Eq.~\eqref{eq:ABSGenericDispersions}, we find $\varepsilon_+\approx t_{34}$ and $\varepsilon_-\approx t_{13}t_{24}/t_{34}$. Thus, there now exists a high- and low-energy sector for the ABSs, with only the $\varepsilon_-$ branch exhibiting a substantial dispersion with $\Delta\phi$, since it reads:
\bea
\varepsilon_-\propto t_{13}t_{24}\propto \cos\chi+\cos\big(\Delta\phi+\varphi\big)\,.
\label{eq:ABSDispersion}
\eea

\begin{figure}[t!]
\centering
\includegraphics[width=1\columnwidth]{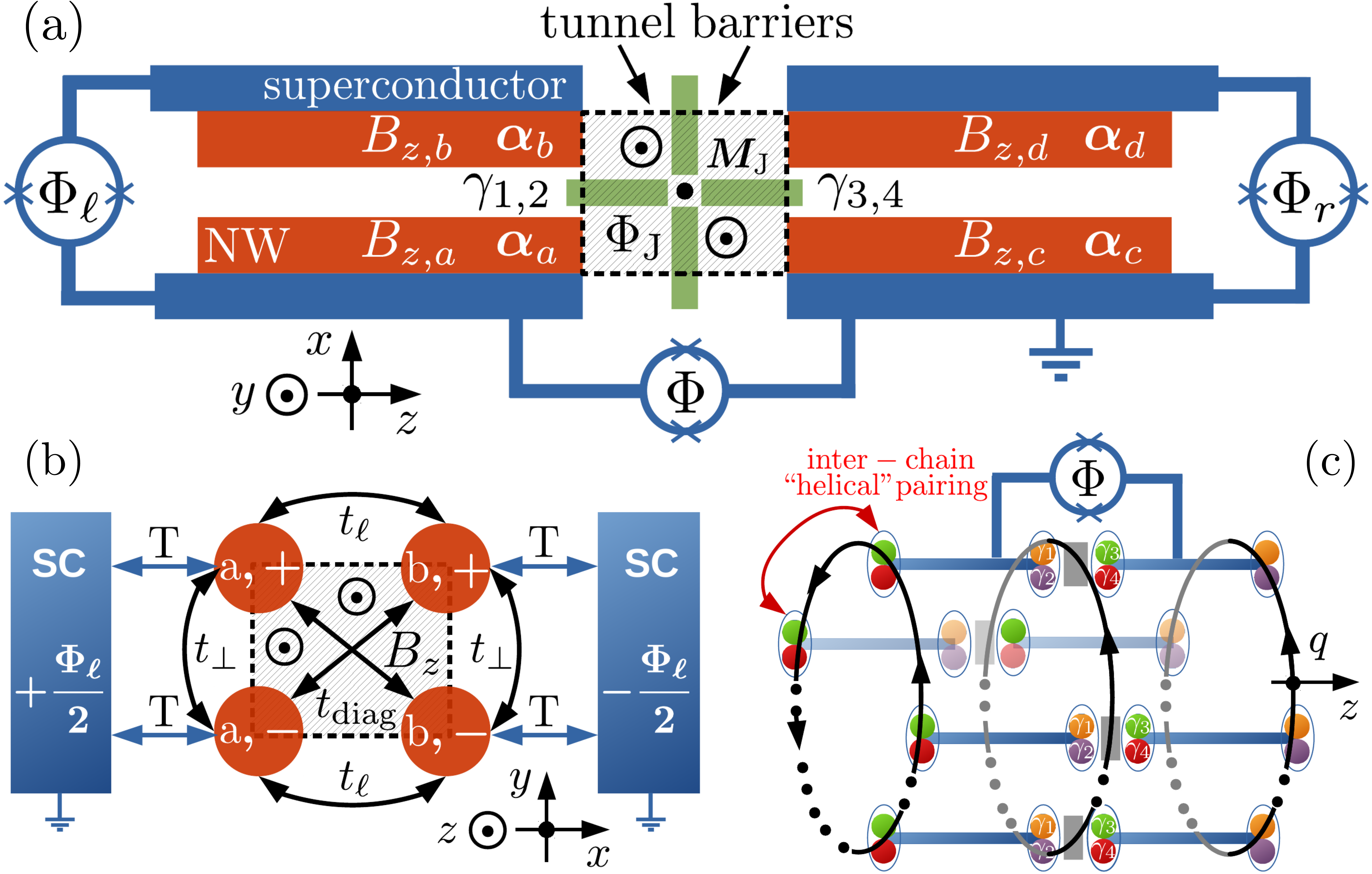}
\protect\caption{(a) Junction of two TSCs, each one consisting of two coupled single-channel semiconductor nanowires (NWs) in pro\-xi\-mi\-ty to conventional SCs. The NWs feel a spin-orbit coupling $(\bm{\alpha}\times\bm{\sigma})_z\hat{p}_z$. For specific values of fluxes $\Phi_{\ell/r}$ and magnetic fields $B_z$, a pair of MBSs appears on each side of the junction. Each pair is protected by a Kramers degeneracy or a sublattice symmetry, which is violated away from the sweet spot. (b) Cross-section of the lhs of a device, which is obtained by stacking two junctions shown in (a). (c) ``Infinite'' 1D network of identical TSC junctions described by the wave number $q$. FP-pumping occurs by varying $\Delta\phi=\Phi$ and an additional parameter, which in: (a) is the interface flux $\Phi_{\rm J}$ (magnetization $\bm{M}_{\rm J}$), (b) is the flux $\sim B_z$ piercing the cross-section, and in (c) is the phase of the inter-chain pairing~\cite{Sup}.}
\label{Fig:Fig3}
\end{figure}

For $\varphi=\chi=0$, Eq.~\eqref{eq:ABSDispersion} yields the quadratic cros\-sing at $\Delta\phi=\pi$, shown in Fig.~\ref{Fig:Fig2}(c), which can be viewed as two merged li\-near crossings of an ABS dispersion as the one in Fig.~\ref{Fig:Fig2}(b). Ano\-ther phase-offset option is $\varphi=\chi=\pi/2$, where one obtains a $\sin(\Delta\phi)$ term, with two linear crossings at $0$ and $\pi$, cf. Fig.~\ref{Fig:Fig2}(d). Additional details are given in Ref.~\onlinecite{JointMercaldo} and the Supplemental Material (SM)~\cite{Sup}. In the former, we show that all the dispersions of Fig.~\ref{Fig:Fig2} are accessible in a TSC junction of two spin-triplet p-wave SCs in the presence of Zeeman fields and spin anisotropy. As we find there, case (c) of Fig.~\ref{Fig:Fig2} appears when the uncoupled interface MBSs are protected by either a chiral or a unitary symmetry, while case (d) arises when the two TSCs possess different chiral symmetries. 

In Fig.~\ref{Fig:Fig3} and the SM~\cite{Sup}, we discuss alternative platforms prominent for ABS engineering. Fig.~\ref{Fig:Fig3}(a) depicts two coupled single-channel semiconductor nanowires (NWs) in pro\-xi\-mi\-ty to conventional SCs on each side of the junction. In addition, magnetic fields ($B_{z,1,2,3,4}$), spin-orbit coupling fields ($\bm{\alpha}_{1,2,3,4}$), and fluxes ($\Phi_{\ell,r}$ and $\Phi=\Delta\phi$) are present. Here, we express the fluxes in terms of the phase differences they induce. This versatile system allows for multiple MBSs per TSC edge protected by a chiral symmetry. The latter is asso\-cia\-ted with either a Kramers degene\-ra\-cy~\cite{Keselman,Jens,KlinovajaTRIparafermions,KlinovajaKramersPairs,Haim,HaimReview} ($\Phi_{\ell,r}=\pi$ and $B_{z,\ell,r}=0$) or a sublattice symmetry~\cite{Jay,KotetesNoZeeman,Hell,PientkaPlanar} ($\Phi_{\ell,r}=0$). The remaining unspecified pa\-ra\-me\-ters are sui\-ta\-bly tuned within a window that allows all four NWs to be in the TSC phase simultaneously. Instead, the slightest de\-via\-tions of the phases and magnetic fields from the values indicated above, violate chiral symmetry and generate the couplings $t_{12,34}$. See also the SM~\cite{Sup}.

Our results open perspectives for nontrivial topo\-lo\-gy in synthetic spaces, when the ABS spectrum is fully-gapped. This becomes possible: \bt{(i)} in devices such as the one in Fig.~\ref{Fig:Fig3}(a) when all MBS couplings are controllable and the interface is flux- or spin-active, \bt{(ii)} in two 1D TSC junctions of Fig.~\ref{Fig:Fig3}(a), stacked as in Fig.~\ref{Fig:Fig3}(b), with each junction supporting gapless ABS spectra, and \bt{(iii)} in infinite networks of identical junctions as in Fig.~\ref{Fig:Fig3}(c). For fully-gapped ABS spectra stemming from four coupled MBSs, one can decompose $i\hat{B}$, which defines a $\mathfrak{so}(4)$ spectrum-generating algebra, into two $\mathfrak{so}(3)$ subalgebras, i.e. $i\hat{B}=\sum_{n=1,2}\bm{g}_n\cdot\bm{L}_n$. Using the Pauli matrices $\bm{\lambda}$ and $\bm{\kappa}$, the $\mathfrak{so}(3)$ generators read $\bm{L}_1=\big(\lambda_1\kappa_2,\,\lambda_2,\,\lambda_3\kappa_2\big)/2$ and $\bm{L}_2=\big(\lambda_2\kappa_1,\,\kappa_2,\,\lambda_2\kappa_3\big)/2$. Therefore, $\bm{g}_1=-\big(t_{13}-t_{24},\,t_{14}+t_{23},\,t_{12}+t_{34}\big)$ and $\bm{g}_2=-\big(t_{13}+t_{24},\,t_{12}-t_{34},\,t_{14}-t_{23}\big)$. See also Ref.~\onlinecite{MeyerHouzet}.

$\mathbb{Z}$-FP pumping and chiral anomaly effects occur when the MBS couplings depend on an additional $2\pi$-periodic variable $\theta$. In this case, one introduces the Berry curvature $\Omega_{\Delta\phi,\theta}^{-,n}$~\cite{Niu} of the occupied eigenstates $\left|u_{-,n}(\Delta\phi,\theta)\right>$ of each $\mathfrak{so}(3)$ subalgebra with $n=1,2$. Swee\-ping $\Delta\phi$ and $\theta$ over an area ${\cal A}:[\Delta\phi_i,\Delta\phi_f]\times[\theta_i,\theta_f]$ yields the pumped charge per period $2\pi$: $\Delta Q({\cal A})=2e\iint_{\cal A}d\Delta\phi d\theta\ph\Omega_{\Delta\phi,\theta}^{-,1}/(2\pi)$~\cite{Sup}. Here, we assumed that $|\bm{g}_1|>|\bm{g}_2|$, other\-wi\-se the Berry curvature $\Omega_{\Delta\phi,\theta}^{-,2}$ should appear above~\cite{Sup}. Moreover, we considered a two-step adiabatic process, in which, the sweeping rates satisfy $\dot{\Delta\phi}\ll\dot{\theta}$, where $\dot{f}\equiv df/dt$. For $|\bm{g}_1|\rightarrow0$, the $\Omega_{\Delta\phi,\theta}^{-,1}$ peaks at the positions of the Berry monopoles, i.e. Weyl points in $(\Delta\phi,\theta,m)$ space, with $m$ an additional pa\-ra\-me\-ter controlling $|\bm{g}_1|$. In the cases of relevance, each Weyl point contributes with $\pm\frac{1}{2}$ to the integral. When ${\cal A}\equiv\mathbb{T}^2=[0,2\pi]\times[0,2\pi]$, we find $\Delta Q/(2e)=C_1\in\mathbb{Z}$, with $C_1$ the respective 1st Chern number~\cite{Niu}. This implies that FP-pumping is not possible, unless \textit{selective swee\-ping} is employed, so that only an odd number number of Weyl points are inside ${\cal A}$ for $|\bm{g}_1|\rightarrow0$.

In Fig.~\ref{Fig:Fig3}(b) we find $\bm{g}_1=\left(m\cos\theta,m\sin\theta,\varepsilon(\Delta\phi)\right)$, with $\varepsilon(\Delta\phi)$ an ABS dispersion with FP-protected crossings at $\Delta\phi=\Delta\phi_c$, as in Figs.~\ref{Fig:Fig2}(b) and~(d). Viewing the phase dif\-fe\-ren\-ce $\Delta\phi$ as a synthetic momentum, yields a Dirac Hamiltonian $\bm{g}_1\cdot\bm{\kappa}$ defined in 2D $(\Delta\phi,t)$ spacetime about each crossing point since, there, $\varepsilon(\Delta\phi)\approx v_c(\Delta\phi-\Delta\phi_c)$ with $v_c$ the slope of the dispersion eva\-lua\-ted at $\Delta\phi=\Delta\phi_c$. Now, Berry monopoles (nodes) appear in $(\Delta\phi,m)$ space at $(\Delta\phi_c,m=0)$, and lead to chiral anomaly effects~\cite{CA}. In the standard 2D Dirac theory, chiral anomaly ma\-ni\-fests as the nonconservation of the chiral charge $\rho_5\propto\left<\kappa_z\right>$ and current $j_5\propto\left<\mathds{1}\right>$ densities. Consequently, the Goldstone-Wilczek formula~\cite{GW} implies that spatial and temporal variations of $\theta$ induce electric charge $\rho\propto\left<\mathds{1}\right>$ and current $j\propto\left<\kappa_z\right>$ densities. Here, the occurence of chiral anomaly implies:
\bea
\Delta Q({\cal A})=-e\frac{\Delta\theta}{\pi}\sum_c{\rm sgn}(v_c)\int_{\cal C}d\varepsilon\ph\delta(\varepsilon-\varepsilon_c)\,,\label{eq:SupercurrentTopo}
\eea

\noi where $\Delta\theta=\theta_f-\theta_i$ and ${\cal C}$ denotes the ABS-dispersion path $\varepsilon_i\mapsto\varepsilon_f$, set by $\Delta\phi_i\mapsto\Delta\phi_f$. The integral is nonzero only if the path includes crossing points of $\varepsilon(\Delta\phi)$. The contribution of each crossing point is ${\rm sgn}(v_c)1$. Thus, for ABS dispersions containing two crossings with opposite slopes, $\Delta Q$ is zero when the path includes both of them, or equi\-va\-len\-tly a quadratic cros\-sing. Notably, FP pumping requires $\Delta\theta=\pi$. Si\-mi\-lar effects emerge for loosely-coupled 1D junctions as in Fig.~\ref{Fig:Fig3}(c), with intrinsic or effective helical-like inter-chain p-wave pairing ${\cal D}(q)\sim qe^{-i\theta}$ being a prerequisite~\cite{JointMercaldo}, since $m\sim q$. The pumped charge \textit{per mode} $q$ is half of that in Eq.~\ref{eq:SupercurrentTopo}~\cite{Sup}, thus, FP-pumping requires $\Delta\theta=2\pi$.

Concluding, we remark that our predictions appear experimentally feasible in organic SCs~\cite{Salts}, purple bronze~\cite{purple}, and hybrid devices~\cite{Fabrizio,Tonio,Ren,Nanocrosses,Krizek}, as explained in the SM~\cite{Sup}. The demonstration of nonstandard ABS dispersions promises to open novel paths for the detection of chiral anomaly and FP-protected cros\-sings, thus, pro\-vi\-ding indirect evidence for MBSs. Finally, the Berry-monopole tomography proposed here is robust against disorder and parameter detunings, as long as these only modify the locations of the monopoles.

\section{Acknowledgements}

PK would like to thank Kasper Grove Rasmussen and Michele Burrello for helpful discussions.


\appendix

\begin{widetext}

\section*{{\Large Supplemental Material}}

\begin{figure*}[b!]
\centering
\includegraphics[width=0.65\textwidth]{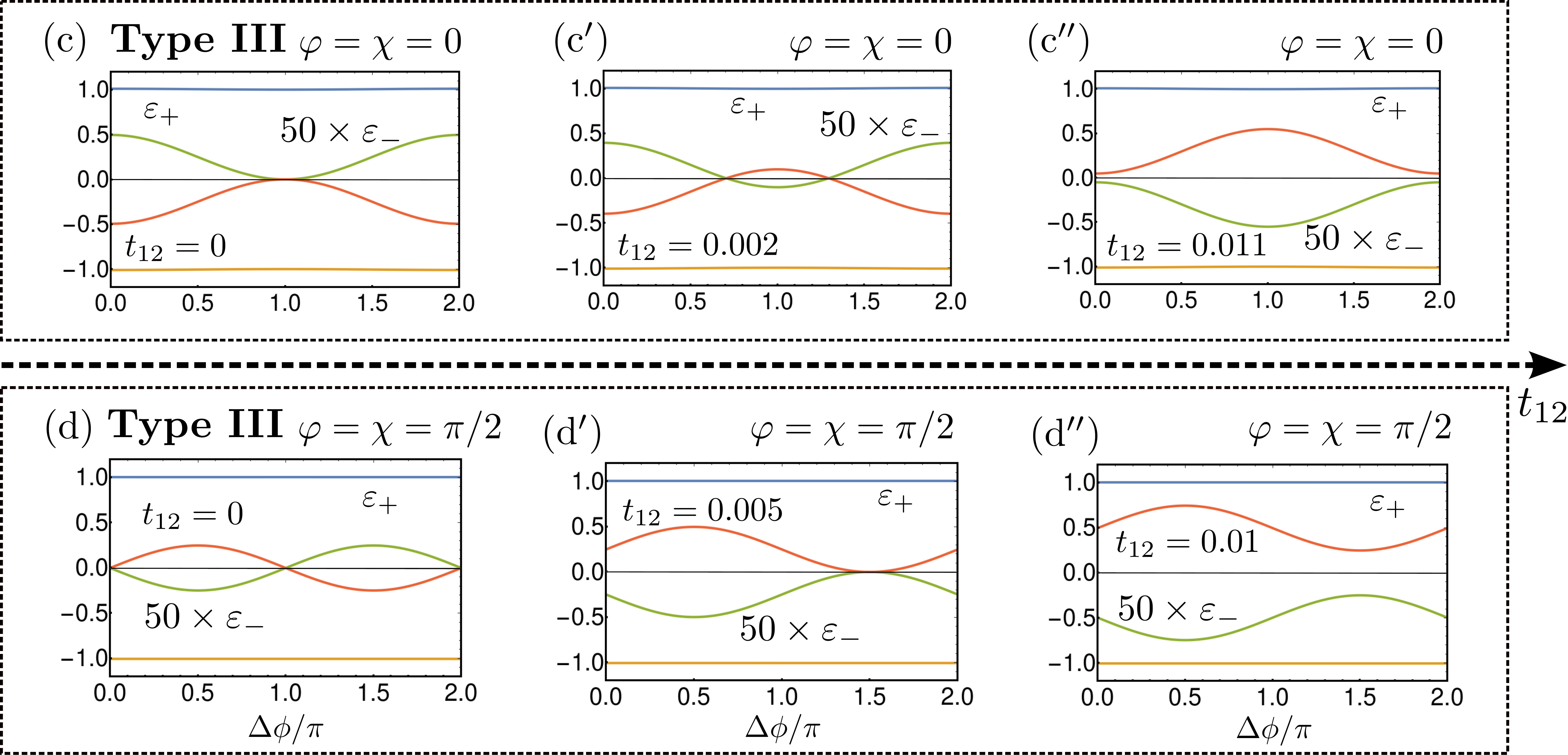}
\protect\caption{Evolution of the Andreev-bound-state (ABS) spectra shown in panels (c) and (d) of Fig.~2 (main text). Here, panels (c)-(c$'$)-(c$''$) and (d)-(d$'$)-(d$''$) portray the evolutions of cases (c) and (d) of the manuscript, upon varying $t_{12}$. As $t_{12}$ decreases, i.e., (c$''$)$\rightarrow$(c$'$)$\rightarrow$(c), the chiral or unitary symmetry becomes restored in one of the topological superconductors (TSCs) and the two linear crossings of (c$''$) merge into the quadratic crossing of (c). In (d)$\rightarrow$(d$'$)$\rightarrow$(d$''$), we depict how a sinusoidal dispersion arises. One observes that the dispersion has a sinusoidal character already before the crossing, hence, manifestly implying that the crossing is fermion-parity (FP) protected. In the above, we employed the same parameter values as in Fig.~2 of the manuscript.}
\label{Fig:FigSup}
\end{figure*}

\section{I. Evolution and Inter-Relation of Andreev-Bound-State Bandstructures}

In this paragraph we present the evolution of the bandstructures of panels (c) and (d) in Fig.~2 of the main text, when $t_{12}$ is varied. Regarding panel (c), one finds that the sequence in Figs.~\ref{Fig:FigSup}(c)-(c$'$)-(c$''$) shown below is accessed. Indeed, star\-ting with $t_{12}\gg t_{13}^{(0)}, t_{24}^{(0)}$, we find that lowe\-ring the strength of $t_{12}$ allows us to sequentially access cases (c$''$)$\rightarrow$(c$'$)$\rightarrow$(c) which have the same structure as the cases shown in panels (a)$\rightarrow$(b)$\rightarrow$(c) of Fig.~2 in the manuscript. In Fig.~\ref{Fig:FigSup}(d)-(d$'$)-(d$''$) depicted below, we exemplify how the band (d) of Fig.~2 (main text) de\-ve\-lops from a situation analogous to Fig.~2(a) (main text).

\section{II. Topological Superconductor Consisting of Two Coupled Semiconductor Nanowires in proximity to a conventional superconductor}\label{sec:Hamiltonian}

Here, we analyze the topological properties of one of the two hybrid devices appearing on a given side of the junction in Fig.~3(a) of the main text. We assume that the strength of the chiral-symmetry violating terms, as well as that for electron tunneling across the junction are much smaller than the energy scale set by the bulk energy gap of each hybrid device. Therefore, one can first study the devices on each side separately, and then, consider their coupling at the level of a low-energy Hamiltonian involving the arising interface Majorana bound states (MBSs).

Each one of these two hybrid devices consists of two tunnel-coupled single-channel semiconductor nanowires (NWs) in proximity to a conventional superconductor (SC) whose segments are characterized by the same superfluid density but generally a different phase. The SC segments are grounded and act as particle reservoirs for the NWs, thus, setting the values $\mu_{a,b,c,d}$ for their chemical potentials~\cite{Mikkelsen,Antipov}. The $s$-th NW is considered to be under the influence of structural and/or bulk  inversion asymmetry generating the spin-orbit coupling term $\big(\bm{\alpha}_s\times\bm{\sigma}\big)_z\hat{p}_z$~\cite{Winkler}. Additional external fields appear in Fig.~3(a). The first type of these consists of the fluxes $\Phi_{\ell/r}$ which impose a phase bias between the two SC segments of a given side of the junction, as well as the flux $\Phi$. These fluxes induce the phase differences $\Delta\phi_{\ell,r}$ and $\Delta\phi$, respectively, and are supposed to be threaded through DC (or RF depending on the implementation) SQUIDs (superconducting quantum interference devices) that connect two successive SC segments. Note that in some cases, we use the fluxes and their corresponding phase differences interchangeably. The last ingredient of the $s$-th NW is an applied magnetic/exchange field $B_{z,s}$ along the $z$ direction and is only present in the NW region. The magnetic field induces spin-splitting by means of the Zeeman effect which can drive the topological phase transition. Given the low-dimensional character of the structure, the orbital effects of the field can be neglected for this setup. The Hamiltonian describing one of these hybrid devices, \textit{say the one on the lhs}, consists of the following three parts, i.e. ${\cal H}_{\ell}={\cal H}_{\rm SC}+{\cal H}_{{\rm NW},\ell}+{\cal H}_{{\rm T},\ell}$:
\bea
{\cal H}_{\rm SC}&=&\int d\bm{r}\ph\bigg\{\sum_{\alpha=\uparrow,\downarrow}
c_{\alpha}^{\dag}(\bm{r})\left(\frac{\hat{\bm{p}}^2}{2m}-\mu\right)c_{\alpha}(\bm{r})+
\Delta\left[e^{i\phi(\bm{r})}c_{\uparrow}^{\dag}(\bm{r})c_{\downarrow}^{\dag}(\bm{r})+{\rm h.c.}\right]\bigg\}\,,\\
{\cal H}_{{\rm NW},\ell}&=&\int_{-\infty}^{+\infty}dk_z
\sum_{\alpha=\uparrow,\downarrow}\bigg\{\sum_{\beta=\uparrow,\downarrow}
\sum_{s=a,b}\psi_{\alpha,s}^{\dag}(k_z)\bigg[\left(\frac{\hbar^2k_z^2}{2m^*}-\mu_{{\rm NW},s}\right)\mathds{1}^{\alpha\beta}\no\\
&&+\hbar k_z\left(\alpha_{y,s}\sigma_x^{\alpha\beta}-\alpha_{x,s}\sigma_y^{\alpha\beta}\right)-B_{z,s}\sigma_z^{\alpha\beta}\bigg]\psi_{\beta,s}(k_z)
+t_\ell\left[\psi_{\alpha,a}^{\dag}(k_z)\psi_{\alpha,b}(k_z)+{\rm h.c.}\right]\bigg\}\,,\\
{\cal H}_{{\rm T},\ell}&=&
\int_{-\infty}^{+\infty}dz\sum_{\alpha=\uparrow,\downarrow}\sum_{s=a,b}{\rm T}_s\bigg[c_{\alpha}^{\dag}(x=x_s,y=0,z)\psi_{\alpha,s}(z)+{\rm h.c.}\bigg]\,,
\eea

\noi with $\bm{\sigma}$ denoting the spin Pauli matrices and $t_{\ell}$ defining the inter-NW tunnel coupling. The chemical potentials $\mu_{{\rm NW},s}$ characterize the NWs before contact to the SC segments. Further, $m$ ($m^*$) denotes the (effective) electron mass in the SC (NW). In the above, we introduced the creation operator $c_{\alpha}^{\dag}(\bm{r})$ of an electron in the SC at position $\bm{r}$ and the creation operator $\psi_{\alpha,s}^{\dag}(z)$ ($\psi_{\alpha,s}^{\dag}(k_z)$) of an electron in the $s$-th NW and position $z$ (wave number $k_z$). In both cases, the index $\alpha$ denotes the spin projection $\alpha=\uparrow,\downarrow$. For the SC-NW coupling we have considered local tunnel couplings ${\rm T}_s$ at the NW positions $x_s$, thus, here not allowing for the possibility of crossed Andreev reflection. One integrates out the electrons of the SCs and obtains a self-energy for the NW electrons, which for energies much smaller than the SC bulk gap $|\Delta|$, leads to the following approximate form for the induced pairing Hamiltonian (in analogy to Ref.~\onlinecite{Potter}): 
\bea
{\cal H}_{\rm pairing}\approx\int_{-\infty}^{+\infty}dk_z\sum_{s=a,b}
\Gamma_s\bigg[e^{i\phi_s}\psi_{\uparrow,s}^{\dag}(k_z)\psi_{\downarrow,s}^{\dag}(-k_z)+{\rm h.c.}\bigg]\,,
\eea

\noi with $\Gamma_s$ the strength of each proximity-induced pairing gap. In this limit, we do not include any renormalization effects other than shifting the bare chemical potentials $\mu_{{\rm NW},s}\mapsto \mu_{s}$, thus, reflecting the electrostatic contact to the SC segments. See also Ref.~\cite{Mikkelsen,Antipov,Potter}. The phases can be rewritten as $\phi_a=\phi_{\ell}+\Delta\phi_{\ell}/2$ and $\phi_b=\phi_{\ell}-\Delta\phi_{\ell}/2$, with $\phi_{\ell}$ the global SC phase of the device on the lhs of the junction, and $\Delta\phi_{\ell}=\phi_a-\phi_b$ the phase difference of the two SC segments on the lhs of the junction, which is set by the respective flux $\Phi_\ell$. 

After introducing the Pauli matrices $\bm{\tau}$ and $\bm{\kappa}$ acting in Nambu (electron-hole) and NW ($\{a,b\}$) spaces respectively, we obtain the following Bogoliubov - de Gennes (BdG) Hamiltonian: 
\bea
\hat{{\cal H}}_{{\rm BdG},\ell}(k_z)&=&\left(\frac{\hbar^2k_z^2}{2m^*}-\mu_{\ell}\right)\tau_z
-\big(B_{z,\ell}+\delta B_{z,\ell}\kappa_z\big)\tau_z\sigma_z+\hbar k_z\big(\alpha_{y,\ell}+\delta \alpha_{y,\ell}\kappa_z\big)\sigma_x-
\hbar k_z\big(\alpha_{x,\ell}+\delta \alpha_{x,\ell}\kappa_z\big)\tau_z\sigma_y\no\\
&+&t_\ell\tau_z\kappa_x-\delta \mu_\ell\tau_z\kappa_z-e^{i\phi_\ell\tau_z}e^{i\Delta\phi_\ell\tau_z\kappa_z/2}\tau_y\big(\Gamma_\ell+\delta\Gamma_\ell\kappa_z\big)\sigma_y
\label{eq:BdGHamiltonian}
\eea

\noi where we employed the notation ${\cal O}_{\ell}=({\cal O}_{a}+{\cal O}_{b})/2$, ${\cal O}_{r}=({\cal O}_{c}+{\cal O}_{d})/2$, $\delta{\cal O}_{\ell}=({\cal O}_{a}-{\cal O}_{b})/2$, $\delta{\cal O}_{r}=({\cal O}_{c}-{\cal O}_{d})/2$, and introduced the spinor:
\bea
\Psi^{\dag}_{\ell}(k_z)=\left(\psi_{\uparrow,a}^{\dag}(k_z),\,\psi_{\downarrow,a}^{\dag}(k_z),\,\psi_{\uparrow,b}^{\dag}(k_z),\,\psi_{\downarrow,b}^{\dag}(k_z),\,
\psi_{\uparrow,a}(-k_z),\,\psi_{\downarrow,a}(-k_z),\,\psi_{\uparrow,b}(-k_z),\,\psi_{\downarrow,b}(-k_z)\right)
\eea

\noi so that ${\cal H}_{{\rm NW},\ell}^{\rm eff}=\frac{1}{2}\int_{-\infty}^{+\infty}dk_z\Psi^{\dag}_{\ell}(k_z)\hat{{\cal H}}_{{\rm BdG},\ell}(k_z)\Psi_{\ell}(k_z)$. The BdG Hamiltonian of Eq.~\eqref{eq:BdGHamiltonian} possesses a charge-conjugation symmetry with operator $\hat{\Xi}=\tau_x\hat{{\cal K}}$, where $\hat{{\cal K}}$ defines the complex-conjugation operator. To proceed we gauge away the global phase factor from the pairing term, since this only affects the tunnel coupling across the junction and enters in $\Delta\phi=\phi_\ell-\phi_r$. Therefore, we obtain the Hamiltonian: 
\bea
\hat{{\cal H}}_{{\rm BdG},\ell}'(k_z)&=&\left(\frac{\hbar^2k_z^2}{2m^*}-\mu_{\ell}\right)\tau_z
-\big(B_{z,\ell}+\delta B_{z,\ell}\kappa_z\big)\tau_z\sigma_z+\hbar k_z\big(\alpha_{y,\ell}+\delta \alpha_{y,\ell}\kappa_z\big)\sigma_x-
\hbar k_z\big(\alpha_{x,\ell}+\delta \alpha_{x,\ell}\kappa_z\big)\tau_z\sigma_y\no\\
&+&t_\ell\tau_z\kappa_x-\delta \mu_\ell\tau_z\kappa_z-e^{i\Delta\phi_\ell\tau_z\kappa_z/2}\tau_y\big(\Gamma_\ell+\delta\Gamma_\ell\kappa_z\big)\sigma_y\,.
\label{eq:BdGHamiltonianNewGauge}
\eea

\noi Depending on the values of the several fields and device parameters, one finds the following topological scenarios:

\subsection{A. Topological Scenario of Kramers Degeneracy}

In this situation, we consider $B_{z,a,b}=0$ and $\Delta\phi_{\ell}=\pi$. Hence, the ensuing Hamiltonian reads: 
\bea
\hat{{\cal H}}_{\rm BdG,\ell}'(k_z)&=&\left(\frac{\hbar^2k_z^2}{2m^*}-\mu_{\ell}\right)\tau_z
+\hbar k_z\big(\alpha_{y,\ell}+\delta \alpha_{y,\ell}\kappa_z\big)\sigma_x-\hbar k_z\big(\alpha_{x,\ell}+\delta \alpha_{x,\ell}\kappa_z\big)\tau_z\sigma_y\no\\
&+&t_\ell\tau_z\kappa_x-\delta \mu_\ell\tau_z\kappa_z-\tau_x\big(\Gamma_\ell\kappa_z+\delta\Gamma_\ell\big)\sigma_y
\label{eq:BdGHamiltonianNewGaugeKramers}
\eea

\noi and possesses additional symmetries, which consist of the standard time-reversal symmetry and a chiral symmetry. Given the particular choice of the spinor, the former is effected by the operator $\hat{\Theta}=i\tau_z\sigma_y\hat{{\cal K}}$, and the latter by $\hat{\Pi}=\tau_y\sigma_y$. Thus, the Hamiltonian belongs to the DIII symmetry class which supports a $\mathbb{Z}_2$ topological invariant in 1D~\cite{Altland,KitaevClassi,Ryu}. This is obtained by introducing the anti-symmetric sewing matrix $\hat{W}(k_{\cal I})=\tau_z\sigma_y\hat{{\cal H}}_{\rm BdG,\ell}'(k_{\cal I})$, defined at the inversion-symmetric wave numbers satisfying $k_z\equiv-k_z$. There exist two such wave numbers in this 1D continuum description, i.e. $k_z=0$ and $k_z=+\infty\equiv-\infty$. The $\mathbb{Z}_2$ invariant takes the values $\pm1$ and its sign is determined by the sign of the product of the Pfaffians of $\hat{W}$ evaluated at all $k_{\cal I}$. Since there can be no gap closing at $|k_z|=\infty$, the sign of the $\mathbb{Z}_2$ invariant is governed by the gap closing at $k_z=0$, which is given by the condition:
\bea
\bigg[t_\ell^2-\left(\mu_{\ell}^2-\delta\mu_\ell^2\right)-\left(\Gamma_\ell^2-\delta\Gamma_\ell^2\right)\bigg]^2+
4\big(\mu_{\ell}\delta\Gamma_\ell-\Gamma_\ell\delta\mu_\ell\big)^2=0\,.
\eea

\noi The above implies that there can be a gap closing when $|t_\ell|<\sqrt{\mu_\ell^2-\delta\mu_\ell^2+\Gamma_\ell^2-\delta\Gamma_\ell^2}$, where we assumed that $|\delta \mu_\ell|<|\mu_\ell|$ and $|\delta \Gamma_\ell|<|\Gamma_\ell|$. To understand the mechanisms that drive the topological phase transitions better, we consider for convenience that $\delta\Gamma_{\ell}=\delta\mu_\ell=\alpha_{y,\ell}=\alpha_{x,\ell}=0$. In this case we have:
\bea
\hat{{\cal H}}_{\rm BdG,\ell}'(k_z)=e^{i\eta_\ell\tau_z\sigma_z/2}\left[\left(\frac{\hbar^2k_z^2}{2m^*}-\mu_{\ell}\right)\tau_z+\delta\alpha_{\ell}\hbar k_z\kappa_z\sigma_x
+t_\ell\tau_z\kappa_x-\Gamma_\ell\tau_x\kappa_z\sigma_y\right]e^{-i\eta_\ell\tau_z\sigma_z/2}
\label{eq:BdGHamiltonianNewGaugeKramersKeselman}
\eea

\noi where we introduced $\delta \alpha_{y,\ell}=\delta \alpha_\ell\cos\eta_\ell$ and $\delta \alpha_{x,\ell}=\delta \alpha_\ell\sin\eta_\ell$. We observe that the two spin-orbit coupling terms can be simplified by performing a rotation in spin space about the $z$ axis. The resulting phase factor can be gauged away and only considered when discussing the Josephson junction, since the TSCs on the lhs and rhs may be characterized by a different orientation angles $\eta_{\ell,r}$ of the spin-orbit vectors in the $xy$ plane.

The above model was first discussed by the authors of Ref.~\cite{Keselman}. In this form, it becomes apparent that the inter-NW hopping term plays a similar role to the one of the Zeeman field in the prototypical single-channel NW TSC model~\cite{LutchynPRL,OregPRL}, albeit the difference it leads to a splitting in NW instead of spin space. This allows for the appearance of MBS Kramers pairs when the condition $|t_\ell|>\sqrt{\mu_{\ell}^2+\Gamma_\ell^2}$ is satisfied. Note that in the present case it is crucial that the Rashba coefficients of the two NWs are opposite, i.e. $\alpha_{y,a}\alpha_{y,b}<0$ and $\alpha_{x,a}\alpha_{x,b}<0$, so that the system has access to the topologically-nontrivial regime. 

The Hamiltonian inside the brackets of Eq.~\eqref{eq:BdGHamiltonianNewGaugeKramersKeselman}, that we denote $\hat{{\cal H}}_{\rm BdG,\ell}''(k_z)$ from now on, possesses a unitary symmetry ${\cal O}=\tau_z\sigma_x$, which implies the presence of an additional set of time-reversal, chiral and charge-conjugation symmetries effected by the respective operators $\hat{\Theta}'=\sigma_z\hat{{\cal K}}$, $\hat{\Pi}'=\tau_x\sigma_z$ and $\hat{\Xi}'=i\tau_y\sigma_x\hat{{\cal K}}$. The combined presence of time-reversal and charge-conjugation operators squaring to $\pm1$, does not allow us to perform a symmetry classification of this Hamiltonian. One is required to block-diagonalize the Hamiltonian with blocks belonging to irreducible representations of the unitary symmetry operator ${\cal O}$. Cf. Refs.~\onlinecite{Ryu,KotetesClassi}. Therefore, we block-diagonalize ${\cal O}$ by performing the unitary transformation ${\cal U}=(\sigma_z+\tau_z\sigma_x)/\sqrt{2}$ and find two block Hamiltonians labelled by $\sigma=\pm1$ corresponding to the eigenstates of $\sigma_z$:
\bea
\underline{\hat{{\cal H}}}_{\rm BdG,\ell,\sigma}''(k_z)=\tau_z\left[\left(\frac{\hbar^2k_z^2}{2m^*}-\mu_{\ell}\right)+
\sigma \delta\alpha_{\ell}\hbar k_z\kappa_z+t_\ell\kappa_x\right]+\Gamma_\ell\tau_y\kappa_z\,.
\label{eq:BdGHamiltonianNewGaugeKramersKeselmanRotated}
\eea

\noi The above block Hamiltonians belong to the BDI symmetry class which supports a $\mathbb{Z}$ topological invariant in 1D given by a winding number~\cite{Ryu}. The state vectors of the resulting MBS Kramers pairs are Kronecker pro\-ducts of the eigenstates of $\tau_x$ and $\sigma_z$, with opposite signs in both spaces. That is, if the state vector for the spin up MBS reads $\left|1(z)\right>=f_1(z)\left|\tau_x=\pm1,\sigma_z=+1\right>$, then the state vector for the spin down state vector is $\left|2(z)\right>=f_2(z)\left|\tau_x=\mp1,\sigma_z=-1\right>$. Here, $f_{1,2}(z)$ are suitable functions decaying away from the interface. We note that in the rotated space the two chiral-symmetry operators become identified with $\tau_x$ and $\tau_x\sigma_z$. Thus, $\left|1(z)\right>$ and $\left|2(z)\right>$ constitute eigenstates of $\tau_x\sigma_z$ with the \textit{same} chirality.

The MBSs Kramers partners are expected to hybridize when $B_{z,a,b}\neq0$ or/and $\Delta\phi_\ell\neq\pi$. In this case, switching on the magnetic fields as well as considering small deviations of the phase difference $\Delta\phi_{\ell}$ from $\pi$, yields the approximate (due to small $\Delta\phi_\ell-\pi$) Hamiltonian:
\bea
\underline{\hat{{\cal H}}}_{\rm BdG,\ell,sym-vio}''(k_z)\approx
\frac{\Delta\phi_{\ell}-\pi}{2}\tau_x\big(\Gamma_\ell+\delta\Gamma_\ell\kappa_z\big)-\big(B_{z,\ell}+\delta B_{z,\ell}\kappa_z\big)\sigma_x\,.
\label{eq:BdGHamiltonianTRSViolating}
\eea

\noi One finds that for the here-considered structure of the spin-orbit coupling, a magnetic field pointing in the $z$ axis leaves the two MBS uncoupled, which is in agreement with Ref.~\onlinecite{Keselman}. This can be understood by identifying the symmetry properties of the Hamiltonian inside the brackets of Eq.~\eqref{eq:BdGHamiltonianNewGaugeKramersKeselman} at the Kramers degeneracy spot, after switching on the field $B_z$. One finds that the system resides in the BDI symmetry class with a chiral symmetry generated by $\tau_x\sigma_z$, and a generalized time-reversal symmetry effected by $\sigma_z\hat{\cal K}$. Since the BDI class supports a $\mathbb{Z}$ invariant, this implies that switching on the magnetic field does not necessarily couple the MBS Kramers pairs, at least for sufficiently weak field-strength. In fact, this holds for all magnetic fields with orientation perpendicular to the spin-orbit coupling vector. Instead, when $\Delta\phi_\ell\neq\pi$, the MBS Kramers partners hybridize even for infinitesimal deviations of $\Delta\phi_\ell$ from $\pi$. Finally, as it was also shown in Ref.~\onlinecite{Keselman}, the violation of Kramers degeneracy opens perspectives for FP by varying $\Delta\phi_\ell$ by $2\pi$. In this case, the pump is constructed as a $\mathbb{Z}_2$ index by virtue of the Kramers degeneracy at $\Delta\phi_\ell=\pi$. 

\subsection{B. Topological Scenario of Sublattice Symmetry}\label{sec:spinless}

Here we consider: $\delta\alpha_{x,\ell}=\delta\alpha_{y,\ell}=0$ and $\Delta\phi_\ell=0$. In this case, the Hamiltonian reads:
\bea
\hat{{\cal H}}_{\rm BdG,\ell}'(k_z)&=&e^{i\eta_\ell\tau_z\sigma_z/2}
\bigg[\left(\frac{\hbar^2k_z^2}{2m^*}-\mu_{\ell}\right)\tau_z-\tau_z\big(B_{z,\ell}+\delta B_{z,\ell}\kappa_z\big)\sigma_z
+\alpha_\ell\hbar k_z\sigma_x\no\\
&+&t_\ell\tau_z\kappa_x-\delta\mu_\ell\tau_z\kappa_z-\tau_y\big(\Gamma_\ell+\delta\Gamma_\ell\kappa_z\big)\sigma_y\bigg]e^{-i\eta_\ell\tau_z\sigma_z/2}\,,
\label{eq:BdGHamiltonianNewGaugeSublattice}
\eea

\noi where we introduced $\alpha_{y,\ell}=\alpha_\ell\cos\eta_\ell$ and $\alpha_{x,\ell}=\alpha_\ell\sin\eta_\ell$. Once again the phase factor related to $\eta_\ell$ can be gauged away and only considered when discussing the coupling of the two TSCs. After removing the phase factors, the Hamiltoninan in brackets denoted $\hat{{\cal H}}_{\rm BdG,\ell}''(k_z)$ possesses additional symmetries to the already existing charge conjugation, and it belongs to the BDI symmetry class which supports a $\mathbb{Z}$ topological invariant in 1D. These symmetries consist of a generalized time-reversal symmetry effected by $\hat{\Theta}=\tau_z\sigma_z\hat{{\cal K}}$ and the chiral symmetry with operator $\hat{\Pi}=\tau_y\sigma_z$. As a consequence, depending on the parameter values, the hybrid system harbors up to two MBSs $\gamma_{1,2}$ near the lhs of the junction, whose state vectors are eigenstates of $\hat{\Pi}$ with the same chirality. To gain further insight, it is helpful to examine the spinless limit, which is reached for a sufficiently strong $B_{z,\ell}$ while $|\delta B_{z,\ell}|\ll|B_{z,\ell}|$. In this limit, one can consider the spin-flip terms inside the brackets of Eq.~\eqref{eq:BdGHamiltonianNewGaugeSublattice} perturbatively~\cite{AliceaTQC}, and obtain an effective Hamiltonian for one of the two spin species. Here,we assume $B_{z,\ell}>0$ without loss of generality, and find the effective Hamiltonian for the spin up electrons where $\sigma_z=+1$
\bea
\hat{{\cal H}}_{\rm BdG,\ell}^{'',\uparrow,{\rm eff}}(k_z)&\approx&
\tau_z\left[\left(\frac{1}{2m^*}-\frac{\alpha_\ell^2}{2B_{z,\ell}}\right)(\hbar k_z)^2
-\left(\mu_{\ell}-\frac{\Gamma_\ell^2+\delta \Gamma_\ell^2}{2B_{z,\ell}}\right)
-\big(B_{z,\ell}+\delta B_{z,\ell}\kappa_z\big)\right]\no\\
&+&t_\ell\tau_z\kappa_x-\left(\delta\mu_\ell-\frac{\Gamma_\ell\delta\Gamma_\ell}{B_{z,\ell}}\right)\tau_z\kappa_z
+\frac{\alpha_\ell}{B_{z,\ell}}\hbar k_z\tau_x\big(\Gamma_\ell+\delta\Gamma_\ell\kappa_z\big)\,,
\label{eq:BdGHamiltonianSpinless}
\eea

\noi In this limit, the chiral-symmetry operator becomes $\hat{\Pi}=\tau_y$, a form that is readily obtainable from the initial expression after setting $\sigma_z=1$. Therefore, in the regime where the above Hamiltonian supports two MBSs per edge, the MBSs state vectors of a pair are characterized by a common eigenvalue of $\tau_y$. Further, their composition in $\kappa$ space is determined by the variables denoting the mismatches of the various quantities between the two NWs, as well as the inter-NW hopping. Assuming that the two NWs are identical, the eigenstates in $\kappa$ space are given by the eigenstates of the operator $\kappa_x$. Finally, note that in the spinless limit, the phase factor ${\rm Exp}(i\eta_\ell\tau_z\sigma_z/2)$ becomes ${\rm Exp}(i\eta_\ell\tau_z/2)$, i.e. the phase related to the orientation of the electric field contributes to the global SC phase and $\phi_\ell\mapsto\phi_\ell+\eta_\ell$.

The sublattice symmetry is violated when $\Delta\phi_\ell\neq0$, in which case the system preserves only the charge-conjugation symmetry and, thus, the Hamiltonian transits to the symmetry class D. Projecting once again Eq.~\eqref{eq:BdGHamiltonianNewGauge} onto the spin-up band and transferring back to coordinate space, yields the symmetry-violating bulk Hamiltonian for small $\Delta\phi_\ell$:
\bea
\hat{{\cal H}}_{\rm BdG,\ell,sym-vio}^{',\uparrow,{\rm eff}}(\hat{p}_z)\approx
-\frac{\Delta\phi_\ell}{2}\frac{\alpha_\ell}{B_{z,\ell}}\hat{p}_z\tau_y\big(\Gamma_\ell\kappa_z+\delta\Gamma_\ell\big)\,.
\label{eq:BdGHamiltonianSpinlessBreakingBad}
\eea

\noi One finds that the term $\propto \Gamma_\ell$ leads to the mixing of the MBSs on a given edge. 

\section{III. Properties of a Josephson Junction of Two Topological Superconductors}

We now explore the properties of a Josephson junction consisting of two TSCs such as the ones studied in the paragraph above. We consider three possible scenarios depending on the type of symmetry protecting the pair of MBSs appearing on each side of the junction. We note that we examine similar possibilities for intrinsic spin-triplet p-wave SCs in our accompanying work of Ref.~\onlinecite{JointMercaldo}. Since we are mainly interested in symmetry aspects and qualitative effects, we can consider for simplicity that the junction is described by a spinful electronic degree of freedom $d_{\uparrow,\downarrow}$ which feels the presence of a magnetization field $\bm{M}=M(\cos\theta_M,\sin\theta_M,0)$, without assuming any interactions at this level of description. This degree of freedom is tunnel-coupled to the NW electrons on both sides of the junction. Under these conditions, the coupling Hamiltonian obtains the general form:
\bea
{\cal H}_{\psi d}=\int dz\sum_{s=a,b,c,d}\sum_{\alpha=\uparrow,\downarrow}
\bigg[\psi_{\alpha,s}^{\dag}(z)e^{-i\phi_s/2}e^{-i\frac{e}{\hbar}\int_{\bm{R}_d}^{\bm{R}_s}\bm{A}\cdot d\bm{r}}\ph{\rm T}_{s}^{\psi d}(z)d_\alpha+{\rm h.c.}\bigg]+
\sum_{\alpha,\beta=\uparrow,\downarrow}d_{\alpha}^{\dag}\bigg(\varepsilon_d\mathds{1}^{\alpha\beta}+\bm{M}\cdot\bm{\sigma}^{\alpha\beta}\bigg)d_{\beta}\,,
\eea

\noi where we additionally considered the presence of a flux $\Phi_{\rm J}$ piercing the interface. The flux is introduced through the orbital coupling to the $y$ component of the magnetic field generated by the vector potential $\bm{A}$. This appears in the tunnel-coupling terms ${\rm T}_{s}^{\psi d}(z)$ through the Peierls substitution~\cite{Peierls}, i.e. by introducing phase factors which are proportional to the line integrals of the vector potential $\bm{A}$. The positions $\bm{R}_s$ denote the terminating points of the NWs, since tunneling is assumed to involve NW electrons which are located only very near the edges. The coordinate vector $\bm{R}_d$ denotes the position of the dot. Here, we assume for simplicity that the positions given by $\bm{R}_{a,b,c,d}$ form a square of an area $S$. 

Assuming that the occupation of the junction degree of freedom does not change, i.e. $|\varepsilon_d|$ is here larger than any relevant ABS energy scale, we can integrate the dot degree of freedom out, and obtain an effective coupling between the two TSCs:
\begin{align}
{\cal H}_{\rm TSC-TSC}=\int dz\int dz'\sum_{s,s'=a,b,c,d}\sum_{\alpha,\beta=\uparrow,\downarrow}
\psi_{\alpha,s}^{\dag}(z)e^{-i\left(\phi_s-\phi_{s'}\right)/2}e^{-i\frac{e}{\hbar}\int_{\bm{R}_{s'}}^{\bm{R}_s}\bm{A}\cdot d\bm{r}}
\bigg[t_{ss'}(z,z')\mathds{1}^{\alpha\beta}+\bm{t}_{ss'}(z,z')\cdot\bm{\sigma}^{\alpha\beta}\bigg]\psi_{\beta,s'}(z')\,,
\end{align}

\noi with the tunnel-coupling functions $t_{ss'}(z,z')$ and $\bm{t}_{ss'}(z,z')$ being real and approximately proportional to $T_{s}^{\psi d}(z)T_{s'}^{\psi d}(z')/|\varepsilon_d|$ and $T_{s}^{\psi d}(z)T_{s'}^{\psi d}(z')\bm{M}/|\varepsilon_d|^2$, respectively. Hence, $\bm{t}_{ss'}(z,z')=|\bm{t}_{ss'}(z,z')|(\cos\theta_M,\sin\theta_M,0)$. Note that the above Hamiltonian also couples NWs on the same side of the junction, thus, it also yields symmetry-violating terms which couple the MBSs on a given side. Therefore, we separate the above Hamiltonian into the following intra-TSC components:
\bea
{\cal H}_{\rm intra-NW,\ell}&=&\frac{1}{2}\int dz\int dz'\ph\Psi_{\ell}^{\dag}(z)
\ph\tau_z\kappa_x\bigg\{t_{ab}(z,z')\mathds{1}+|\bm{t}_{ab}(z,z')|\sigma_xe^{i\theta_M\tau_z\sigma_z}\bigg\}e^{-i\theta_{\rm J}\tau_z\kappa_z}\ph\Psi_{\ell}(z')\,,\\\no\\
{\cal H}_{{\rm intra-NW},r}&=&\frac{1}{2}\int dz\int dz'\ph\Psi_{r}^{\dag}(z)
\ph\tau_z\kappa_x\bigg\{t_{cd}(z,z')\mathds{1}+|\bm{t}_{cd}(z,z')|\sigma_xe^{i\theta_M\tau_z\sigma_z}\bigg\}e^{+i\theta_{\rm J}\tau_z\kappa_z}\ph\Psi_{r}(z')
\eea

\noi and inter-TSC components:
\bea
&&{\cal H}_{\rm inter-TSC}=\frac{1}{2}\int dz\int dz'\ph\Psi_{\ell}^{\dag}(z)e^{-i\Delta\phi\tau_z/2}\bigg\{\no\\
&&\tau_z\left(\begin{array}{cc}
\bigg\{t_{ac}(z,z')\mathds{1}+|\bm{t}_{ac}(z,z')|\sigma_xe^{i\theta_M\tau_z\sigma_z}\bigg\}e^{-i\theta_{\rm J}\tau_z}&
t_{ad}(z,z')\mathds{1}+|\bm{t}_{ad}(z,z')|\sigma_xe^{i\theta_M\tau_z\sigma_z}\\\\
t_{bc}(z,z')\mathds{1}+|\bm{t}_{bc}(z,z')|\sigma_xe^{i\theta_M\tau_z\sigma_z}
&\bigg\{t_{bd}(z,z')\mathds{1}+|\bm{t}_{bd}(z,z')|\sigma_xe^{i\theta_M\tau_z\sigma_z}\bigg\}e^{+i\theta_{\rm J}\tau_z}\end{array}\right)\bigg\}\Psi_{r}(z')+{\rm h.c.},
\no\\\
\eea

\noi where we introduced the multicomponent spinors defined on each side of the junction. The $2\times2$ matrix presented above explicitly, is defined in NW $\kappa$ space. Note that we introduced the phase $\theta_{\rm J}=\pi\nu_{\rm J}/2$, with $\nu_{\rm J}=B_{{\rm J},y}S/(h/e)$ the number of flux quanta piercing the junction's interface area. To obtain this result, we employed the symmetric gauge for the vector potential, i.e. $\bm{A}=B_{{\rm J},y}(z/2,0,-x/2)$. Since we are interested in the low-energy regime and the ABS properties, in the following, we only account for the MBS contribution to the spinors $\Psi_{\ell,r}(z)$.

\subsection{A. MBSs protected by a Kramers degeneracy on both sides of the junction}

To facilitate the calculations, we consider the limit $|t_{\ell,r}|\gg\sqrt{\mu_{\ell,r}^2+\Gamma_{\ell,r}^2}$, so that we can project onto a given eigenstate of the terms $t_{\ell,r}\tau_z\kappa_x$ in $\kappa$ space, i.e. $\kappa_x=+1$ or $\kappa_x=-1$. In analogy to the approach followed for deriving the spinless model of Sec.~\ref{sec:spinless}, we assume that $t_{\ell,r}<0$, and using Eq.~\eqref{eq:BdGHamiltonianNewGaugeKramersKeselmanRotated} we find:
\bea
\underline{\hat{{\cal H}}}_{\rm BdG,\ell,\sigma}^{'',+,{\rm eff}}(k_z)=\tau_z\left[\left(\frac{1}{2m^*}-\frac{\delta\alpha_\ell^2}{2|t_\ell|}\right)(\hbar k_z)^2
-\left(\mu_\ell-\frac{\Gamma_\ell^2}{2|t_\ell|}\right)-|t_\ell|\right]+\frac{\delta\alpha_{\ell}\Gamma_\ell}{|t_\ell|}\sigma\hbar k_z\tau_y\,.
\label{eq:BdGHamiltonianNewGaugeKramersKeselmanRotatedNWless}
\eea

\noi The state vectors of the MBSs are eigenstates of the chiral symmetry operator $\tau_x\sigma_z$. We assume without loss of generality that the MBSs on the lhs of the junction are eigenstates of $\tau_x\sigma_z$ with chirality $+1$. After restoring previously gauged away phase factors, the MBS state vectors on the lhs read in $\tau\otimes\kappa\otimes\sigma$ space:
\bea
\left|1(z)\right>=f_1(z)e^{+i\eta_\ell\tau_z/2}
\frac{1}{2}\left(\begin{array}{c}1\\1\end{array}\right)\otimes
\left(\begin{array}{c}1\\1\end{array}\right)\otimes
\left(\begin{array}{c}1\\0\end{array}\right)
\phd{\rm and}\phd
\left|2(z)\right>=f_2(z)e^{-i\eta_\ell\tau_z/2}
\frac{1}{2}\left(\begin{array}{c}i\\-i\end{array}\right)\otimes
\left(\begin{array}{c}1\\1\end{array}\right)\otimes
\left(\begin{array}{c}0\\1\end{array}\right)\,,
\eea

\noi where $f_{1,2}(z)$ properly normalized functions decaying away from the interface. We proceed with obtaining the MBS couplings. For this we set $\theta_{\rm J}=0$ since, while it can serve as a means to modify the various couplings, it is not an essential ingredient in the particular situation. We consider the two possibilities: 
\begin{enumerate}
\item $\Pi_r=-\Pi_\ell$:
\bea
\left|3(z)\right>=f_3(z)e^{+i\eta_r\tau_z/2}
\frac{1}{2}\left(\begin{array}{c}i\\-i\end{array}\right)\otimes
\left(\begin{array}{c}1\\1\end{array}\right)\otimes
\left(\begin{array}{c}1\\0\end{array}\right)
\phd{\rm and}\phd
\left|4(z)\right>=f_4(z)e^{-i\eta_r\tau_z/2}
\frac{1}{2}\left(\begin{array}{c}1\\1\end{array}\right)\otimes
\left(\begin{array}{c}1\\1\end{array}\right)\otimes
\left(\begin{array}{c}0\\1\end{array}\right)\,.
\eea

We find the following MBS coupling Hamiltonian:
\bea
{\cal H}_{\rm intra-TSC}&=&i\cos\left(\theta_M+\eta_\ell\right)|\bm{t}_{ab}|\gamma_1\gamma_2-i\cos\left(\theta_M+\eta_r\right)|\bm{t}_{cd}|\gamma_3\gamma_4\,,\\\no\\
{\cal H}_{\rm inter-TSC}&=&it_{\ell r}\left[\cos\left(\frac{\Delta\phi+\eta_\ell-\eta_r}{2}\right)\gamma_1\gamma_3-
\cos\left(\frac{\Delta\phi-\eta_\ell+\eta_r}{2}\right)\gamma_2\gamma_4\right]\no\\
&+&i|\bm{t}_{\ell r}|\left[\cos\left(\frac{\Delta\phi+\eta_\ell+\eta_r+2\theta_M+\pi}{2}\right)\gamma_1\gamma_4-
\cos\left(\frac{\Delta\phi-\eta_\ell-\eta_r-2\theta_M-\pi}{2}\right)\gamma_2\gamma_3\right]\,,\label{eq:MBScouplingKramers}
\eea

\noi with $t_{\ell r}=(t_{ac}+t_{ad}+t_{bc}+t_{bd})/2$ and $|\bm{t}_{\ell r}|=(|\bm{t}_{ac}|+|\bm{t}_{ad}|+|\bm{t}_{bc}|+|\bm{t}_{bd}|)/2$. 

\item $\Pi_r=+\Pi_\ell$:
\bea
\left|3(z)\right>=f_3(z)e^{+i\eta_r\tau_z/2}
\frac{1}{2}\left(\begin{array}{c}1\\1\end{array}\right)\otimes\left(\begin{array}{c}1\\1\end{array}\right)\otimes\left(\begin{array}{c}1\\0\end{array}\right)
\phd{\rm and}\phd
\left|4(z)\right>=f_4(z)e^{-i\eta_r\tau_z/2}
\frac{1}{2}\left(\begin{array}{c}i\\-i\end{array}\right)\otimes\left(\begin{array}{c}1\\1\end{array}\right)\otimes\left(\begin{array}{c}0\\1\end{array}\right)\,.
\eea

We find that the MBS coupling Hamiltonian in the present case, can be obtained from the one inferred for $\Pi_r=-\Pi_\ell$ by effecting the shift: $\eta_r\mapsto\eta_r-\pi$.

\end{enumerate}

\subsection{B. MBSs protected by a sublattice symmetry on both sides of the junction}

In this case it is convenient to consider the spinless limit, where the tunnel couplings $\bm{t}_{ab,cd,ad,bc}(z,z')$ eventually drop out. The MBSs state vectors for NWs with $\delta\mu_{\ell}=\delta B_{z,\ell}=\delta\Gamma_{\ell}=0$ on a given side of the junction are eigenstates of $\kappa_x$, and are written as:
\bea
&&\left|1(z)\right>=f_1(z)
e^{i\eta_\ell\tau_z/2}\left|\Pi_{\ell}\right>\otimes\frac{1}{\sqrt{2}}\left(\begin{array}{c}1\\1\end{array}\right)
\otimes\left(\begin{array}{c}1\\0\end{array}\right),\,\phd
\left|2(z)\right>=f_2(z)e^{i\eta_\ell\tau_z/2}\left|\Pi_{\ell}\right>\otimes\frac{1}{\sqrt{2}}\left(\begin{array}{c}1\\-1\end{array}\right)
\otimes\left(\begin{array}{c}1\\0\end{array}\right),\,\\
&&\left|3(z)\right>=f_3(z)e^{i\eta_r\tau_z/2}\left|\Pi_{r}\right>\otimes\frac{1}{\sqrt{2}}\left(\begin{array}{c}1\\1\end{array}\right)
\otimes\left(\begin{array}{c}1\\0\end{array}\right),\,\phd
\left|4(z)\right>=f_4(z)e^{i\eta_r\tau_z/2}\left|\Pi_{r}\right>\otimes\frac{1}{\sqrt{2}}\left(\begin{array}{c}1\\-1\end{array}\right)
\otimes\left(\begin{array}{c}1\\0\end{array}\right)\,,
\eea

\noi with the chiralities $\Pi_{\ell,r}$ coinciding with a specific eigenvalue of $\tau_y$. To obtain the MBS coupling Hamiltonian, we assume without loss of generality that $\left|\Pi_{\ell}\right>=\left|\tau_y=+1\right>$. For the sake of transparency of the inter-TSC terms, we consider the simplifications $t_{ac}=t_{bd}=t_{||}$ and $t_{bc}=t_{ad}=t_{\rm diag}$, which yield:
\bea
{\cal H}_{\rm intra-TSC}&=&i\sin\theta_{\rm J}\left(t_{cd}\gamma_3\gamma_4-t_{ab}\gamma_1\gamma_2\right)\,,\\\no\\
{\cal H}_{\rm inter-TSC}
&=&i\cos\left(\frac{\Delta\phi+\eta_\ell-\eta_r}{2}\right)
\bigg[\big(t_{||}\cos\theta_{\rm J}+t_{\rm diag}\big)\gamma_1\gamma_3+\big(t_{||}\cos\theta_{\rm J}-t_{\rm diag}\big)\gamma_2\gamma_4\bigg]\no\\
&+&it_{||}\sin\theta_{\rm J}\left[\cos\left(\frac{\Delta\phi+\eta_\ell-\eta_r+\pi}{2}\right)\gamma_1\gamma_4-
\cos\left(\frac{\Delta\phi+\eta_\ell-\eta_r-\pi}{2}\right)\gamma_2\gamma_3\right]\qquad
\eea

\noi where we set $\Pi_r=-\Pi_{\ell}$. The coupling Hamiltonian for $\Pi_r=\Pi_{\ell}$ is obtained by performing the shift $\eta_r\mapsto\eta_r-\pi$ in the expressions of the Hamiltonians above. 

\subsection{C. MBSs protected by a different symmetry on the two sides of the junction}

We now investigate the case, in which, the pair of MBSs on the lhs are protected by a Kramers degeneracy, while the pair of MBSs on the rhs is protected by a sublattice symmetry. Without loss of generality, we assume that the chirality of the MBSs on the lhs is $\Pi_\ell=+1$. For the sake of clarity, we consider the simplifications $t_{ac}=t_{bd}=t_{||}$, $t_{bc}=t_{ad}=t_{\rm diag}$, $|\bm{t}_{ac}|=|\bm{t}_{bd}|=|\bm{t}_{||}|$, and $|\bm{t}_{bc}|=|\bm{t}_{ad}|=|\bm{t}_{\rm diag}|$, which lead to:
\bea
&&{\cal H}_{\rm intra-TSC}=
i\cos\theta_{\rm J}\cos\left(\theta_M+\eta_\ell\right)|\bm{t}_{ab}|\gamma_1\gamma_2+it_{cd}\sin\theta_{\rm J}\gamma_3\gamma_4\,,\\\no\\
&&{\cal H}_{\rm inter-TSC}
=i\cos\left(\frac{\Delta\phi+\eta_\ell-\eta_r+\pi/2}{2}\right)\big(t_{||}\cos\theta_{\rm J}+t_{\rm diag}\big)\gamma_1\gamma_3
-i\cos\left(\frac{\Delta\phi-\eta_\ell-\eta_r-2\theta_M+\pi/2}{2}\right)|\bm{t}_{||}|\sin\theta_{\rm J}\gamma_2\gamma_4\no\\
&&-i\cos\left(\frac{\Delta\phi+\eta_\ell-\eta_r-\pi/2}{2}\right)t_{||}\sin\theta_{\rm J}\gamma_1\gamma_4
-i\cos\left(\frac{\Delta\phi-\eta_\ell-\eta_r-2\theta_M-\pi/2}{2}\right)\big(|\bm{t}_{||}|\cos\theta_{\rm J}+|\bm{t}_{\rm diag}|\big)\gamma_2\gamma_3
\eea

\noi where we set $\Pi_r=-\Pi_{\ell}$. The coupling Hamiltonian for $\Pi_r=\Pi_{\ell}$ is obtained by performing the shift $\eta_r\mapsto\eta_r-\pi$ in the expressions of the Hamiltonians above. 

\section{IV. Topological Properties of Two Stacked TSC Josephson Junctions}\label{sec:Stacked}

In this paragraph, we investigate the properties of the device in Fig.~3(b) of the main text, which can be viewed as two stacked devices of Fig.~3(a). Following the spirit of the previous section, we first study the emergence of MBSs in a given layer on a given side of the junction. Later on we consider the coupling between layers on a given side as well as across the junction, at the level of MBSs. Here, we restrict ourselves to the situation in which the MBSs on each side of the junction are protected by a sublattice symmetry. Based on Fig.~3(b) and the results of the previous sections, we obtain the Hamiltonian for one side of the device, \textit{say the one on the lhs} of the stacked junctions. For convenience, we assume that the \textit{all} NWs feel the same chemical potential, proximity-induced superfluid density, as well as Zeeman coupling to the external magnetic field. On the other hand, we assume that the spin-orbit vectors on each side have the same modulus $\alpha_{\ell,r}>0$, but generally different orientations. Under these conditions, we find the partial Hamiltonians: 
\bea
{\cal H}_{\rm SC}&=&\int d\bm{r}\ph\bigg\{\sum_{\alpha=\uparrow,\downarrow}
c_{\alpha}^{\dag}(\bm{r})\left(\frac{\hat{\bm{p}}^2}{2m}-\mu\right)c_{\alpha}(\bm{r})+
\Delta\left[e^{i\phi(\bm{r})}c_{\uparrow}^{\dag}(\bm{r})c_{\downarrow}^{\dag}(\bm{r})+{\rm h.c.}\right]\bigg\}\,,\\
{\cal H}_{{\rm NW},j,\ell}&=&\int_{-\infty}^{+\infty}dk_z
\sum_{\alpha=\uparrow,\downarrow}\bigg\{\sum_{\beta=\uparrow,\downarrow}
\sum_{s=a,b}\psi_{\alpha,s,j}^{\dag}(k_z)\bigg[\left(\frac{\hbar^2k_z^2}{2m^*}-\mu_{{\rm NW}}\right)\mathds{1}^{\alpha\beta}\no\\
&+&\hbar k_z\left(\alpha_{y,j,\ell}\sigma_x^{\alpha\beta}-\alpha_{x,j,\ell}\sigma_y^{\alpha\beta}\right)-B_z\sigma_z^{\alpha\beta}\bigg]\psi_{\beta,s,j}(k_z)
+t_\ell\left[\psi_{\alpha,a,j}^{\dag}(k_z)\psi_{\alpha,b,j}(k_z)+{\rm h.c.}\right]\bigg\}\,,\qquad\\
{\cal H}_{{\rm NW},\perp,\ell}&=&\int_{-\infty}^{+\infty}dk_z\sum_{\alpha=\uparrow,\downarrow}\bigg\{
t_{\perp,\ell}e^{-i\Phi_z}\psi_{\alpha,a,+}^{\dag}(k_z)\psi_{\alpha,a,-}(k_z)+t_{\perp,\ell}e^{+i\Phi_z}\psi_{\alpha,b,+}^{\dag}(k_z)\psi_{\alpha,b,-}(k_z)\no\\
&&\qquad\qquad\qquad\quad+t_{{\rm diag},\ell}\left[\psi_{\alpha,a,+}^{\dag}(k_z)\psi_{\alpha,b,-}(k_z)+\psi_{\alpha,b,+}^{\dag}(k_z)\psi_{\alpha,a,-}(k_z)\right]+{\rm h.c.}\bigg\}\,,\\
{\cal H}_{{\rm T},\ell}&=&
{\rm T}\int_{-\infty}^{+\infty}dz\sum_{\alpha=\uparrow,\downarrow}\sum_{j=\pm}\sum_{s=a,b}\bigg[c_{\alpha}^{\dag}(x=x_s,y=y_j,z)\psi_{\alpha,s,j}(z)+{\rm h.c.}\bigg]\,,
\eea

\noi with $j=\pm$ labelling the upper/lower layer. In the above, we introduced the phase $\Phi_z=\pi\nu_z$, where $\nu_z$ denotes the number of flux quanta threaded by $B_z$ through the (generally) rectangular cross-section formed by the four NWs on the lhs. The quantum of flux here is $h/e$, while we employed the Landau gauge $A_y(x)=B_zx$. Integrating out the SC degrees of freedom yields the intra- and inter-layer pairing Hamiltonian:
\bea
{\cal H}_{{\rm pairing},j}&\approx&\Gamma_{\ell}\int_{-\infty}^{+\infty}dk_z\sum_{s=a,b}\bigg[e^{i\phi_s}\psi_{\uparrow,s,j}^{\dag}(k_z)\psi_{\downarrow,s,j}^{\dag}(-k_z)
+{\rm h.c.}\bigg]\,,\\
{\cal H}_{{\rm pairing},\perp}&\approx&\Gamma_{\perp,\ell}\int_{-\infty}^{+\infty}dk_z\sum_{s=a,b}\bigg\{
e^{i\phi_s}\left[\psi_{\uparrow,s,+}^{\dag}(k_z)\psi_{\downarrow,s,-}^{\dag}(-k_z)+\psi_{\uparrow,s,-}^{\dag}(k_z)\psi_{\downarrow,s,+}^{\dag}(-k_z)\right]
+{\rm h.c.}\bigg\}\,,
\eea

\noi with $j=\pm$. We remark that the inter-layer pairing Hamiltonian is a result of crossed Andreev reflection. The above was obtained under the same assumptions as considered in Sec.~\ref{sec:Hamiltonian}. We proceed by introducing the spinor:
\bea
&&{\cal X}^{\dag}_{\ell}(k_z)=\bigg(
\psi_{\uparrow,a,+}^{\dag}(k_z),\,\psi_{\downarrow,a,+}^{\dag}(k_z),\,
\psi_{\uparrow,b,+}^{\dag}(k_z),\,\psi_{\downarrow,b,+}^{\dag}(k_z),\,
\psi_{\uparrow,a,-}^{\dag}(k_z),\,\psi_{\downarrow,a,-}^{\dag}(k_z),\,
\psi_{\uparrow,b,-}^{\dag}(k_z),\,\psi_{\downarrow,b,-}^{\dag}(k_z),\,\no\\
&&\qquad
\psi_{\uparrow,a,+}(-k_z),\,\psi_{\downarrow,a,+}(-k_z),\,
\psi_{\uparrow,b,+}(-k_z),\,\psi_{\downarrow,b,+}(-k_z),\,
\psi_{\uparrow,a,-}(-k_z),\,\psi_{\downarrow,a,-}(-k_z),\,
\psi_{\uparrow,b,-}(-k_z),\,\psi_{\downarrow,b,-}(-k_z)\bigg)\qquad
\eea

\noi so that ${\cal H}_{{\rm NW},\ell}^{\rm eff}=\frac{1}{2}\int_{-\infty}^{+\infty}dk_z\ph{\cal X}^{\dag}_{\ell}(k_z)\hat{{\cal H}}_{{\rm BdG},\ell}(k_z){\cal X}_{\ell}(k_z)$, with:
\bea
\hat{{\cal H}}_{{\rm BdG},\ell}(k_z)&=&e^{i\phi_\ell\tau_z/2}
\bigg\{\left(\frac{\hbar^2k_z^2}{2m^*}-\mu_{\ell}\right)\tau_z-B_z\tau_z\sigma_z+
\alpha_{\ell}\hbar k_z\left(\frac{1+\lambda_z}{2}e^{i\eta_{+,\ell}\tau_z\sigma_z}+\frac{1-\lambda_z}{2}e^{i\eta_{-,\ell}\tau_z\sigma_z}\right)\sigma_x
+t_\ell\tau_z\kappa_x\no\\
&-&\Gamma_\ell e^{i\Delta\phi_\ell\tau_z\kappa_z/2}\tau_y\sigma_y+t_{\perp,\ell}e^{-i\Phi_z\tau_z\lambda_z\kappa_z}\tau_z\lambda_x+t_{{\rm diag},\ell}\tau_z\lambda_x\kappa_x
-\Gamma_{\perp,\ell}e^{i\Delta\phi_\ell\tau_z\kappa_z/2}\tau_y\lambda_x\sigma_y\bigg\}e^{-i\phi_\ell\tau_z/2}\,,
\label{eq:BdGHamiltonianStackedNewGauge}
\eea

\noi with $\alpha_{y,\pm,\ell}=\alpha_{\ell}\cos\eta_{\pm,\ell}$ and $\alpha_{x,\pm,\ell}=\alpha_{\ell}\sin\eta_{\pm,\ell}$. In addition, we introduced the $\bm{\kappa}$ and $\bm{\lambda}$ Pauli matrices acting in intra-layer $\{a,b\}$ and inter-layer $\{+,-\}$ spaces, respectively. For the remainder, we consider the spinless limit with $B_z>0$. Based on the same method followed in Sec.~\ref{sec:spinless}, we find the effective intra- and inter-layer Hamiltonians for the spin-up band:
\bea
\hat{{\cal H}}_{{\rm BdG},j,\ell}^{\uparrow,{\rm eff}}(k_z)&\approx&e^{i(\phi_\ell+\eta_\ell)\tau_z/2}
\bigg\{
\left[\left(\frac{1}{2m^*}-\frac{\alpha_\ell^2}{2B_z}\right)(\hbar k_z)^2-\left(\mu_{\ell}-\frac{\Gamma_\ell^2+\Gamma_{\perp,\ell}^2}{2B_z}\right)-B_z\right]\tau_z
+t_{\ell}\tau_z\kappa_x\no\\
&&+\frac{\alpha_\ell\Gamma_\ell}{B_z}\hbar k_ze^{i\Delta\phi_\ell\tau_z\kappa_z/2}e^{ji\Delta\eta_\ell\tau_z/2}\tau_x\bigg\}e^{-i(\phi_\ell+\eta_\ell)\tau_z/2}\,,\no\\
\hat{{\cal H}}_{{\rm BdG},\perp,\ell}^{\uparrow,{\rm eff}}(\hat{p}_z)&\approx&e^{i(\phi_\ell+\eta_\ell)\tau_z/2}
\bigg\{
\left(\frac{\Gamma_\ell\Gamma_{\perp,\ell}}{B_z}+t_{\perp,\ell}e^{-i\Phi_z\tau_z\lambda_z\kappa_z}\right)\tau_z\lambda_x
+\frac{\alpha_\ell\Gamma_{\perp,\ell}}{B_z}\hat{p}_z\cos\left(\frac{\Delta\eta_\ell}{2}\right)e^{i\Delta\phi_\ell\tau_z\kappa_z/2}\tau_x\lambda_x\no\\
&&+t_{{\rm diag},\ell}\tau_z\lambda_x\kappa_x\bigg\}e^{-i(\phi_\ell+\eta_\ell)\tau_z/2}\,,
\label{eq:BdGHamiltonianStackedNewGaugeSpinless}
\eea

\noi where $j=\pm$, while we introduced $\eta_\ell=(\eta_{+,\ell}+\eta_{-,\ell})/2$ and $\Delta\eta_\ell=\eta_{+,\ell}-\eta_{-,\ell}$. The inter-layer terms are here assumed to be weak perturbations, i.e. $t_{\perp,\ell}$, $t_{{\rm diag},\ell}$ and $\Gamma_{\perp,\ell}$ lead to terms whose energy scale is much smaller than the bulk energy gap arising from the intra-layer Hamiltonians. For loosely-coupled layers, we account for the inter-layer terms only at the level of the low-energy Hamiltonian involving MBSs. The latter appear when sublattice symmetry is present, i.e. for $\Delta\phi_\ell=0$, and they are characterized by state vectors which consist of a Kronecker product of the chiral-symmetry matrix $\tau_y$ eigenstates, the eigenstates of $\lambda_z$ and the eigenstates of $\kappa_x$. In more detail, one finds the following four MBS state vectors in $\tau\otimes\lambda\otimes\kappa$ space (we omit the spin-space component since we restrict to spin up):
\bea
\left|1,+(z)\right>=\frac{f_1(z)}{\sqrt{2}}e^{+i\Delta\eta_\ell\tau_z/4}\left|\Pi_{\ell}\right>\otimes\left(\begin{array}{c}1\\0\end{array}\right)
\otimes\left(\begin{array}{c}1\\1\end{array}\right),\,&\ph&
\left|2,+(z)\right>=\frac{f_2(z)}{\sqrt{2}}e^{+i\Delta\eta_\ell\tau_z/4}\left|\Pi_{\ell}\right>\otimes\left(\begin{array}{c}1\\0\end{array}\right)
\otimes\left(\begin{array}{c}1\\-1\end{array}\right),\,\no\\
\left|1,-(z)\right>=\frac{f_1(z)}{\sqrt{2}}e^{-i\Delta\eta_\ell\tau_z/4}\left|\Pi_{\ell}\right>\otimes\left(\begin{array}{c}0\\1\end{array}\right)
\otimes\left(\begin{array}{c}1\\1\end{array}\right),\,&\ph&
\left|2,-(z)\right>=\frac{f_2(z)}{\sqrt{2}}e^{-i\Delta\eta_\ell\tau_z/4}\left|\Pi_{\ell}\right>\otimes\left(\begin{array}{c}0\\1\end{array}\right)
\otimes\left(\begin{array}{c}1\\-1\end{array}\right)\,,\qquad
\eea

\begin{figure*}[t!]
\centering
\includegraphics[width=0.91\textwidth]{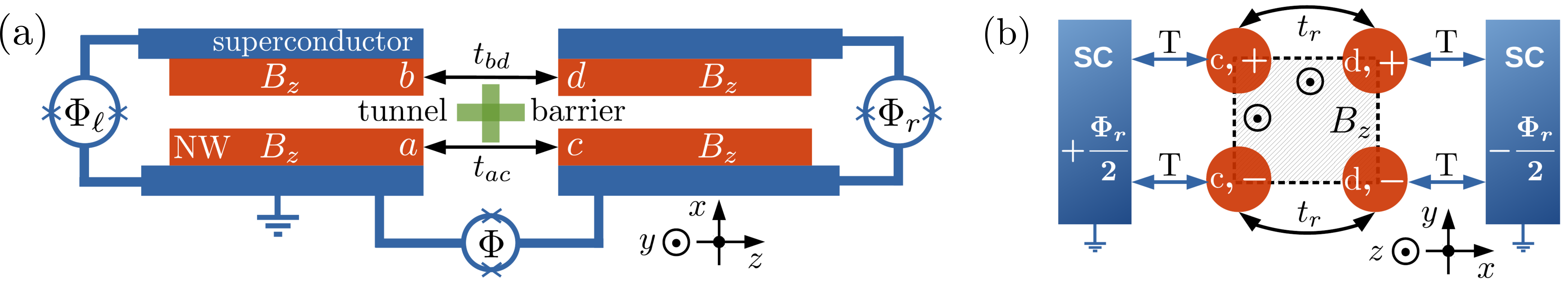}
\protect\caption{(a) Top view of the device discussed in Sec.~\ref{sec:Stacked}, corresponding to Fig.~3(b) in the manuscript. The two layers are not visible here. (b) Cross-section of the layered structure on the rhs of the device. Ideally, the two rhs NW layers are uncoupled.}
\label{Fig:FigSup2}
\end{figure*}

\noi with the chirality $\Pi_\ell$ coinciding with a specific eigenvalue of $\tau_y$. In the above, we have dropped a common phase factor related to the effective global superconducting phase $\phi_\ell+\eta_\ell$ on the left part of the complete device, which is only relevant when considering the MBS couplings across the junction. By taking into account the inter-layer terms, as well as \textit{small} deviations of $\Delta\phi_\ell$ from zero, we find the MBS coupling Hamiltonian (at lowest order in the expansion parameters $t_{\perp,\ell}$, $t_{{\rm diag},\ell}$, $\Gamma_{\perp,\ell}$ and $\Delta\phi_\ell$):
\bea
{\cal H}_{{\rm MBS},\ell}&\approx&i\Pi_\ell\Delta\phi_\ell\frac{\hbar\alpha_\ell\Gamma_\ell}{2\xi_\ell B_z}\big(\gamma_{1+}\gamma_{2+}+\gamma_{1-}\gamma_{2-}\big)
-it_{\perp,\ell}\sin\Phi_z\cos\left(\frac{\Delta\eta_\ell}{2}\right)\big(\gamma_{1+}\gamma_{2-}+\gamma_{2+}\gamma_{1-}\big)\no\\
&-&i\sin\left(\frac{\Delta\eta_\ell}{2}\right)\bigg[\left(t_{\perp,\ell}\cos\Phi_z+\frac{\Gamma_\ell\Gamma_{\perp,\ell}}{B_z}
+t_{{\rm diag},\ell}\right)\gamma_{1+}\gamma_{1-}
+\left(t_{\perp,\ell}\cos\Phi_z+\frac{\Gamma_\ell\Gamma_{\perp,\ell}}{B_z}-t_{{\rm diag},\ell}\right)\gamma_{2+}\gamma_{2-}\bigg]\,,\qquad
\eea

\noi with $1/\xi_\ell=\int dz\ph f_1(z)f_2'(z)$, where $f'=df/dz$. We also approximately set $\int dz\ph f_1(z)f_2(z)\approx1$. 

The Hamiltonian describing the two stacked layers on the rhs is similar to the one for the lhs, after the relabellings $\ell\mapsto r$. The MBS state vectors are also obtained by the $\left|1\pm\right>$ and $\left|2\pm\right>$, after the mappings $\{1,2,\ell\}\mapsto\{3,4,r\}$. In regards with the coupling across the junction, we consider the presence of tunnel barriers which impose that only NWs which are located opposite of the junction are coupled. Further assuming $\Pi_\ell=-\Pi_r=1$, yields the couplings per given layer (for simplicity we set $t_{ac}=t_{bd}=t_{||}$ for both layers): 
\bea
{\cal H}_{{\rm inter-TSC},j}=it_{||}\cos\left(\frac{\Delta\phi+\eta_{j,\ell}-\eta_{j,r}}{2}\right)\big(\gamma_{1,j}\gamma_{3,j}+\gamma_{2,j}\gamma_{4,j}\big)\,.
\eea

Since we are interested in the case where the sublattice symmetry is mainly broken on one of the sides of the junction, \textit{say on the rhs}, we consider the ideal scenario $t_{\perp,r}=t_{{\rm diag},r}=0$ and $\Gamma_{\perp,\ell/r}\approx0$. Here, we considered that there is no tunnel coupling between the two layers on the rhs, as well as we dropped the terms related to crossed Andreev reflection. The latter are not essential to the phenomena of interest and their inclusion only leads to quantitative modifications. Thus, we assume that the distance of the NWs is sufficiently larger than the Fermi wave length in the SC in which case they can be safely ignored. In addition, we consider that the sublattice symmetry is violated only very weakly on the lhs, while the term violating the sublattice symmetry on the rhs constitutes the dominant ABS-level energy scale. Therefore, we obtain the conclusive set of Hamiltonians for the lhs and rhs: 
\bea
{\cal H}_{{\rm MBS},\ell}&=&it_{12}\big(\gamma_{1+}\gamma_{2+}+\gamma_{1-}\gamma_{2-}\big)
-i\cos\left(\frac{\Delta\eta_\ell}{2}\right)t_{\perp,\ell}\sin\Phi_z\big(\gamma_{1+}\gamma_{2-}+\gamma_{2+}\gamma_{1-}\big)\no\\
&-&i\sin\left(\frac{\Delta\eta_\ell}{2}\right)\bigg[\big(t_{\perp,\ell}\cos\Phi_z+t_{{\rm diag},\ell}\big)\gamma_{1+}\gamma_{1-}+
\big(t_{\perp,\ell}\cos\Phi_z-t_{{\rm diag},\ell}\big)\gamma_{2+}\gamma_{2-}\bigg]\,,\\
{\cal H}_{{\rm MBS},r}&=&it_{34}\big(\gamma_{3+}\gamma_{4+}+\gamma_{3-}\gamma_{4-}\big)\,,
\eea

\noi where $t_{12}=\Pi_\ell\Delta\phi_\ell\hbar\alpha_\ell\Gamma_\ell/(2\xi_\ell B_z)$ and $t_{34}=\Pi_r\Delta\phi_r\hbar\alpha_r\Gamma_r/(2\xi_r B_z)$. Since $t_{34}$ defines the largest energy scale, we can integrate the MBS on the rhs out, and obtain a renormalized sublattice-symmetry breaking term for the lhs, i.e.
\bea
{\cal H}_{{\rm MBS},\ell}^{\rm eff}&=&\sum_{j=\pm}\left[t_{12}-\frac{t_{||}^2}{t_{34}}
\cos^2\left(\frac{\Delta\phi+\eta_{j,\ell}-\eta_{j,r}}{2}\right)\right]i\gamma_{1j}\gamma_{2j}
-i\cos\left(\frac{\Delta\eta_\ell}{2}\right)t_{\perp,\ell}\sin\Phi_z\big(\gamma_{1+}\gamma_{2-}+\gamma_{2+}\gamma_{1-}\big)\no\\
&-&i\sin\left(\frac{\Delta\eta_\ell}{2}\right)\bigg[\big(t_{\perp,\ell}\cos\Phi_z+t_{{\rm diag},\ell}\big)\gamma_{1+}\gamma_{1-}+
\big(t_{\perp,\ell}\cos\Phi_z-t_{{\rm diag},\ell}\big)\gamma_{2+}\gamma_{2-}\bigg]\,.\label{eq:EffectiveStacked}
\eea

\section{V. Infinite Networks of 1D TSC Josephson Junctions}

In this section we address the case of ``infinite'' coupled identical chains. Here, we base our analysis on the platform of the previous paragraph, while in Ref.~\onlinecite{JointMercaldo} we investigate a number of scenarios in connection to intrinsic spin-triplet p-wave SCs. We start from the Hamiltonian in Eq.~\eqref{eq:EffectiveStacked}, and write it in a compact form as follows:
\bea
{\cal H}_{{\rm MBS},n}
=i\varepsilon(\Delta\phi)\big(\gamma_{1,n}\gamma_{2,n}+\gamma_{1,n-1}\gamma_{2,n-1}\big)+\frac{i}{2}\bigg[v_1\gamma_{1,n}\gamma_{1,n-1}+v_2\gamma_{2,n}\gamma_{2,n-1}
-u\big(\gamma_{1,n}\gamma_{2,n-1}+\gamma_{2,n}\gamma_{1,n-1}\big)\bigg]\,,
\eea

\noi where $+\mapsto n$ and $-\mapsto n-1$. In this manner, the upper and lower chain becomes the $n$-th and $n-1$-th chain. We note that, since the chains are assumed to be identical, we set $\eta_{+,\ell}-\eta_{+,r}=\eta_{-,\ell}-\eta_{-,r}$, which allowed us to introduce the ABS dispersion $\varepsilon(\Delta\phi)$, which has a similar structure to the dispersions of scenarios II and III of the main text. At this stage, we consider that the system consists of $N$ identical chains and the invariance under translations allows us to introduce the wave number $q$, as well as the corresponding operators:
\bea
\gamma_{s,n}=\frac{1}{\sqrt{N}}\sum_qe^{iqn}\gamma_{s,q}\quad{\rm where}\quad \gamma_{s,q}^{\dag}=\gamma_{s,-q}
\quad{\rm and}\quad \{\gamma_{s,q},\gamma_{s',-q'}\}=\delta_{s,s'}\delta_{q,q'}\quad{\rm with}\quad s=1,2\,.
\eea

\noi Summing over the chains and employing the relations above, lead to the Hamiltonian:
\begin{align}
{\cal H}_{{\rm MBS}}=\frac{1}{2}\sum_q\bigg[
i\varepsilon(\Delta\phi)\big(\gamma_{1,-q}\gamma_{2,q}-\gamma_{2,-q}\gamma_{1,q}\big)+v_1\sin q\gamma_{1,-q}\gamma_{1,q}+v_2\sin q\gamma_{2,-q}\gamma_{2,q}
-u\sin q\big(\gamma_{1,-q}\gamma_{2,q}+\gamma_{2,-q}\gamma_{1,q}\big)\bigg]\,.
\end{align}

\noi To this end, a remark is in place. While the above expression appears to hold $\forall q$, in fact, it holds only in the vicinity of $q=0$ or $q=\pi$. The reason for this is that the Majorana modes $\gamma_{s,q}$ are obtained only near one of the two inversion-symmetric points, since in the limit of loosely-coupled chains considered here, the bulk topological phase transition for a single chain, can only happen at one of there two wave numbers. See also Ref.~\onlinecite{JointMercaldo}. In this manner, we can focus in the vicinity of one of the two points, \textit{say the $q=0$}, and adopt a continuum description, so that $\sum_q\mapsto\int dq$, $\gamma_{s,q}\mapsto\gamma_{s}(q)$ and $\sin q\mapsto q$. By introducing the $\bm{\tau}$ Pauli matrices, we rewrite the Hamiltonian as:
\bea
{\cal H}_{{\rm MBS}}&=&\frac{1}{2}\int dq\left(\begin{array}{cc}\gamma_1(-q)&\gamma_2(-q)\end{array}\right)
\left(\begin{array}{cc}v_1q&-uq+i\varepsilon(\Delta\phi)\\-uq-i\varepsilon(\Delta\phi)&v_2q\end{array}\right)
\left(\begin{array}{c}\gamma_1(q)\\\gamma_2(q)\end{array}\right)\no\\
&=&\frac{1}{2}\int dq\ph\bm{A}^{\dag}(q)\left[{\cal J}(q)\mathds{1}+\bm{g}(\Delta\phi,q)\cdot\bm{\tau}\right]\bm{A}(q)\label{eq:AndreevMode}
\eea

\noi where we introduced the Andreev mode Nambu spinor $\bm{A}^{\dag}(q)=(a^{\dag}(q)\ph a(-q))$, with $a(q)=(\gamma_1(q)+i\gamma_2(q))/\sqrt{2}$. Here, $\{a(q),a^{\dag}(q')\}=\delta(q-q')$ and $\{a(q),a(q')\}=0$. In addition, we set:
\bea
{\cal J}(q)=\frac{v_1+v_2}{2}q\quad{\rm and}\quad \bm{g}(\Delta\phi,q)=\left(\frac{v_1-v_2}{2}q,uq,\varepsilon(\Delta\phi)\right)\,.
\eea

\noi The quantity ${\cal J}(q)$ corresponds to a supercurrent, while the term ${\cal D}(q)=\big[(v_1-v_2)/2-iu\big]q$ corresponds to a p-wave pairing term, since it enters as ${\cal D}(q)a^{\dag}(q)a^{\dag}(-q)+{\rm h.c.}$. The energy spectrum for the Andreev modes reads: $E_{\pm}(\Delta\phi,q)={\cal J}(q)\pm\sqrt{|{\cal D}(q)|^2+\left[\varepsilon(\Delta\phi)\right]^2}$. The above implies that when ${\rm sgn}(v_1v_2)>0$ the pairing is \textit{chiral}, since the  dispersions of the Majorana modes $\gamma_{1,2}(q)$ have the same slope sign. Instead, when ${\rm sgn}(v_1v_2)<0$ the pairing is \textit{helical}, since the dispersions of the Majorana modes $\gamma_{1,2}(q)$ have opposite slope sign. As discussed in the manuscript, favorable conditions for FP pumping and chiral anomaly phenomena appear for a gapped spectrum. Therefore, this becomes possible only in the case in which the chains demonstrate helical pairing. In the situation of the previous paragraph, this implies $|t_{\perp,\ell}\cos\Phi_z|<|t_{{\rm diag},\ell}|$.

\section{VI. Fermion-Parity Pumping: General Theory}

We now proceed with analyzing the technical details underlying the emergence of FP pumping and chiral anomaly discussed in the manuscript. We start from the generic Hamiltonian describing the coupling of four MBSs. As pointed out in the manuscript, when considering fully-gapped spectra, this can be written in the following manner:
\bea
{\cal H}=\frac{i}{2}\bm{\Gamma}^{\intercal}\hat{B}(\Delta\phi,\theta)\bm{\Gamma}=
\frac{1}{2}\bm{\Gamma}^{\intercal}\hat{\cal H}_{\rm MBS}(\Delta\phi,\theta)\bm{\Gamma}=
\frac{1}{2}\sum_{s=\pm}\left(\begin{array}{cc}a_s^{\dag}&a_s\end{array}\right)
\left(\begin{array}{cc}\varepsilon_s(\Delta\phi,\theta)&0\\0&-\varepsilon_s(\Delta\phi,\theta)\end{array}\right)
\left(\begin{array}{cc}a_s\\a_s^{\dag}\end{array}\right)\,,
\eea

\noi where $\hat{\cal H}_{\rm MBS}(\Delta\phi,\theta)=\sum_{n=1,2}\hat{{\cal H}}_n(\Delta\phi,\theta)$ with $\hat{\cal H}_n(\Delta\phi,\theta)=\bm{g}_n(\Delta\phi,\theta)\cdot\bm{L}_n$. Each $\mathfrak{so}(3)$ Hamiltonian labelled by $n=1,2$, is characterized by two eigenstates denoted here $\left|u_{\pm,n}(\Delta\phi,\theta)\right>$ corresponding to eigenvalues $\pm|\bm{g}_n(\Delta\phi,\theta)|/2$. The eigenstates $\left|{\cal U}_{\pm,\pm}\right>$ and respective eigenenergies $E_{\pm,\pm}$ of the total Hamiltonian $\hat{\cal H}_{\rm MBS}(\Delta\phi,\theta)$ are given by:
\bea
\left|{\cal U}_{\pm,s}(\Delta\phi,\theta)\right>=\left|u_{\pm,1}(\Delta\phi,\theta)\right>\otimes\left|u_{\pm s,2}(\Delta\phi,\theta)\right>\,&{\rm with}&\,
E_{\pm,s}(\Delta\phi,\theta)=\pm\varepsilon_s(\Delta\phi,\theta)=\pm\frac{|\bm{g}_1(\Delta\phi,\theta)|+s|\bm{g}_2(\Delta\phi,\theta)|}{2}\qquad
\eea

\noi and $s=\pm$. To proceed, we first identify the current operator for the matrix Hamiltonian $\hat{{\cal H}}_{\rm MBS}$. By definition the current operator can be determined using the response to a probe flux $\Phi_p$ 
\bea
\hat{J}_{\rm sc}(\Delta\phi,\theta)=
-\left.\frac{\partial\hat{{\cal H}}_{\rm MBS}(\Delta\phi+\frac{2e}{\hbar}\Phi_p,\theta)}{\partial\Phi_p}\right|_{\Phi_p=0}
=-\frac{2e}{\hbar}\frac{\partial\hat{\cal H}_{\rm MBS}(\Delta\phi,\theta)}{\partial\Delta\phi}\,,
\eea

\noi with $-e<0$ the electron's charge. At this point, we consider that $\theta$ varies very slowly in time, and $\Delta\phi$ varies even slower than $\theta$, so that it is eligible to apply time-dependent perturbation theory based on the instantaneous eigenstates of the system for a quasi-static $\Delta\phi$ as $\theta$ varies, cf. Ref~\onlinecite{Niu}. For $\theta(t)$ varying in time at a constant rate $\dot{\theta}$, we find:
\bea
\left|{\cal U}_{\nu,s}(\Delta\phi,t)\right>\approx \left|{\cal U}_{\nu,s}(\Delta\phi,\theta)\right>
+\sum_{(\nu',s')\neq(\nu,s)}\frac{i\hbar\dot{\theta}\left|{\cal U}_{\nu',s'}(\Delta\phi,\theta)\right>}
{E_{\nu',s'}(\Delta\phi,\theta)-E_{\nu,s}(\Delta\phi,\theta)}
\left<{\cal U}_{\nu',s'}(\Delta\phi,\theta)\right|\frac{\partial}{\partial\theta}\left|{\cal U}_{\nu,s}(\Delta\phi,\theta)\right>\,,
\eea

\noi with $\nu,s=\pm$. Therefore, the supercurrent $J_{\rm sc}(\Delta\phi)$ flown is given by:
\bea
J_{\rm sc}(\Delta\phi,\theta)=e\sum_{\nu,s=\pm}\left\{
-\left<{\cal U}_{\nu,s}(\Delta\phi,\theta)\right|\frac{1}{\hbar}\frac{\partial\hat{\cal H}_{\rm MBS}(\Delta\phi,\theta)}{\partial\Delta\phi}
\left|{\cal U}_{\nu,s}(\Delta\phi,\theta)\right>
+\dot{\theta}\Omega_{\Delta\phi,\theta}^{\nu,s}(\Delta\phi,\theta)\right\}f(E_{\nu,s})\,,
\eea

\noi where we introduced the Fermi-Dirac distribution $f(\epsilon)$ for a given energy $\epsilon$, and the Berry curvature of the ABS levels $\nu,s=\pm$ in $(\Delta\phi,\theta)$ space: 
\bea
\Omega_{\Delta\phi,\theta}^{\nu,s}(\Delta\phi,\theta)=
i\left\{\left[\frac{\partial}{\partial\Delta\phi}\left<{\cal U}_{\nu,s}(\Delta\phi,\theta)\right|\right]
\left[\frac{\partial}{\partial\theta}\left|{\cal U}_{\nu,s}(\Delta\phi,\theta)\right>\right]-\Delta\phi\leftrightarrow\theta\right\}\,.
\eea

\noi Taking into account of the fact that the eigenstates $\left|{\cal U}_{\nu,s}(\Delta\phi,\theta)\right>$ can be written as Kronecker products of the form $\left|u_{\nu,1}(\Delta\phi,\theta)\right>\otimes\left|u_{\nu*s,2}(\Delta\phi,\theta)\right>$, allows us to express the Berry curvature above as
\bea
\Omega_{\Delta\phi,\theta}^{\nu,s}(\Delta\phi,\theta)=
\Omega_{\Delta\phi,\theta}^{\nu,1}(\Delta\phi,\theta)+\Omega_{\Delta\phi,\theta}^{\nu*s,2}(\Delta\phi,\theta)\,,
\eea

\noi where we introduced the Berry curvatures $\Omega_{\Delta\phi,\theta}^{\nu,n}(\Delta\phi,\theta)$ of the eigenstates $\left|u_{\nu,n}(\Delta\phi,\theta)\right>$ with $n=1,2$. Even more, by making use of the property $\Omega_{\Delta\phi,\theta}^{-\nu,n}(\Delta\phi,\theta)=-\Omega_{\Delta\phi,\theta}^{\nu,n}(\Delta\phi,\theta)$, we find that, if only the two negative ABS branches are occupied, the sum over the Berry curvatures yields:
\bea
\sum_{s=\pm}\Omega_{\Delta\phi,\theta}^{-,s}(\Delta\phi,\theta)=
\sum_{s=\pm}\left[\Omega_{\Delta\phi,\theta}^{-,1}(\Delta\phi,\theta)+\Omega_{\Delta\phi,\theta}^{-s,2}(\Delta\phi,\theta)\right]=
2\Omega_{\Delta\phi,\theta}^{-,1}(\Delta\phi,\theta)\,.
\eea

\noi We note that the above result is obtained under the condition $|\bm{g}_1|>|\bm{g}_2|$, otherwise the roles of $\bm{g}_1$ and $\bm{g}_2$ should be exchanged. Based on the above, the pumped charge when $\theta_i\mapsto\theta_f$ for a fixed $\Delta\phi$ is given by:
\bea
\Delta Q(\Delta\phi,\theta_i\mapsto\theta_f)=\int_{t_i}^{t_f}dt\ph J_{\rm sc}^{\dot{\theta}}(\Delta\phi)
=2e\int_{\theta_i}^{\theta_f}d\theta\ph\Omega_{\Delta\phi,\theta}^{-,1}(\Delta\phi,\theta)\,.
\eea

\noi The above is obtained under the condition that the phase $\Delta\phi$ is varied much slower than $\theta$, i.e. $\dot{\Delta\phi}\ll\dot{\theta}$. Therefore, varying $\Delta\phi_i\mapsto\Delta\phi_f$, one finds the transferred charge per period $2\pi$:
\bea
\Delta Q({\cal A})=2e\iint_{\cal A}\frac{d\Delta\phi d\theta}{2\pi}\ph\Omega_{\Delta\phi,\theta}^{-,1}(\Delta\phi,\theta)\,,
\eea

\noi with ${\cal A}=[\Delta\phi_i,\Delta\phi_f]\times[\theta_i,\theta_f]$. We now examine the outcome of the above expression for a $\bm{g}_1$ vector of the form $\bm{g}_1=(m\cos\theta,m\sin\theta,\varepsilon(\Delta\phi))$. Since the nodes are located on the $\Delta\phi$ axis, one can sweep $\theta$ with no restrictions, since the Berry monopoles do not live in $\theta$ space, and obtain:
\bea
\Delta Q({\cal A})=-e\frac{\Delta\theta}{\pi}\sum_c{\rm sgn}(v_c)\int_{\cal C}d\varepsilon\ph\frac{m^2}{2\sqrt{(\varepsilon-\varepsilon_c)^2+m^2}^3}=
-e\frac{\Delta\theta}{\pi}\sum_c{\rm sgn}(v_c)\int_{\cal C}d\varepsilon\ph\delta(\varepsilon-\varepsilon_c)\,,\label{eq:CApump}
\eea

\noi with $v_c$ given by $d\varepsilon/d\Delta\phi$ evaluated at the dispersion's crossing points $\Delta\phi_c$ satisfying $\varepsilon(\Delta\phi_c)=0$. We find that the above pumped charge can only modify the FP if we pump $\theta$ for half a period, which reflects the requirement for selective parameter sweeping.

\section{VII. Fermion-Parity Pumping: Case Studies}

In this paragraph we briefly examine the emergence of FP pumping in the devices put forward in the main text and the present supplemental file.

\subsection{A. Single TSC Josephson junction with MBSs protected by a Kramers degeneracy on both sides}

For simplicity, we consider the case with $\Delta\phi_\ell=\Delta\phi_r=\pi$ and $|\bm{t}_{ab}|=|\bm{t}_{cd}|=|\bm{t}_{\rm intra}|$. Hence, we find:
\bea
\bm{g}_1&=&\left(
-2t_{\ell r}\cos\left(\frac{\Delta\eta}{2}\right)\cos\left(\frac{\Delta\phi}{2}\right),
2|\bm{t}_{\ell r}|\sin\left(\frac{\Delta\phi}{2}\right)\cos\left(\theta_M+\eta_c\right),
2|\bm{t}_{\rm intra}|\sin\left(\frac{\Delta\eta}{2}\right)\sin\big(\theta_M+\eta_c\big)\right)\,,\quad\\
\bm{g}_2&=&\left(
2t_{\ell r}\sin\left(\frac{\Delta\eta}{2}\right)\sin\left(\frac{\Delta\phi}{2}\right),
-2|\bm{t}_{\rm intra}|\cos\left(\frac{\Delta\eta}{2}\right)\cos\big(\theta_M+\eta_c\big),
2|\bm{t}_{\ell r}|\cos\left(\frac{\Delta\phi}{2}\right)\sin\left(\theta_M+\eta_c\right)\right)\,,
\eea

\noi where $\Delta\eta=\eta_\ell-\eta_r$ and $\eta_c=(\eta_\ell+\eta_r)/2$. We assume that the parameters are tuned so that $|\bm{g}_1|>|\bm{g}_2|$ (without loss of generality). In this case, we focus on $\bm{g}_1$. Assuming that $\Delta\eta\neq0,\pi$ \& $t_{\ell r}\neq0$ \& $|\bm{t}_{\ell r}|\neq0$, the Berry monopoles are found in $(\Delta\phi,\theta_M+\eta_c,|\bm{t}_{\rm intra}|)$ space at the locations $(\pi,\pm\pi/2,0)$.

\subsection{B. Single TSC Josephson junction with MBSs protected by a sublattice symmetry on both sides}

For simplicity, we consider the case with $\Delta\phi_\ell=0$ and find:
\bea
\bm{g}_1&=&\left(-2t_{\rm diag}\cos\left(\frac{\Delta\phi+\Delta\eta}{2}\right),2t_{||}\sin\theta_{\rm J}\sin\left(\frac{\Delta\phi+\Delta\eta}{2}\right),
(t_{ab}-t_{cd})\sin\theta_{\rm J}-\epsilon_{{\rm vio},r}\Delta\phi_r\right)\,,\\
\bm{g}_2&=&\left(-2t_{||}\cos\theta_{\rm J}\cos\left(\frac{\Delta\phi+\Delta\eta}{2}\right),(t_{ab}+t_{cd})\sin\theta_{\rm J}+\epsilon_{{\rm vio},r}\Delta\phi_r,0\right)\,,
\eea

\noi with $\Delta\eta=\eta_\ell-\eta_r$, while the energy scale $\epsilon_{{\rm vio},r}$ can be read off from Eq.~\eqref{eq:BdGHamiltonianSpinlessBreakingBad} after $\ell\mapsto r$. We consider that the parameters are experimentally tuned so that $|\bm{g}_1|>|\bm{g}_2|$. Assuming that $t_{ab}\neq t_{cd}\neq0$ \& $\epsilon_{{\rm vio},r}\neq0$ \& $t_{\rm diag}\neq0$ ($t_{||}\neq0$), the Berry monopoles are found in $(\Delta\phi+\Delta\eta,\theta_{\rm J},t_{||})$ ($(\Delta\phi+\Delta\eta,\theta_{\rm J},t_{\rm diag})$) space, at the locations $(\pi,\theta_{c},0)$ ($(0,\theta_c,0)$), with $(t_{ab}-t_{cd})\sin\theta_c=\epsilon_{{\rm vio},r}\Delta\phi_r$.

\subsection{C. Single TSC Josephson junction with MBSs protected by a different symmetry on each side}

In this case, there is a large number of parameters, and depending on the experimental setup one can obtain various scenarios for the Berry monopoles. Here, we only focus on the special situation of an interface which is only flux-active. For simplicity, we further set $t_{cd}=0$. These yield: 
\bea
\bm{g}_1&=&\left(-(t_{||}\cos\theta_{\rm J}+t_{\rm diag})\cos\left(\frac{\Delta\phi+\Delta\eta+\pi/2}{2}\right),
t_{||}\sin\theta_{\rm J}\cos\left(\frac{\Delta\phi+\Delta\eta-\pi/2}{2}\right),-\frac{\Delta\phi_\ell-\pi}{2}\Gamma_\ell-\epsilon_{{\rm vio},r}\Delta\phi_r\right)\,,\\
\bm{g}_2&=&\left(-(t_{||}\cos\theta_{\rm J}+t_{\rm diag})\cos\left(\frac{\Delta\phi+\Delta\eta+\pi/2}{2}\right),-\frac{\Delta\phi_\ell-\pi}{2}\Gamma_\ell+\epsilon_{{\rm vio},r}\Delta\phi_r,t_{||}\sin\theta_{\rm J}\cos\left(\frac{\Delta\phi+\Delta\eta-\pi/2}{2}\right)\right)\,.\qquad
\eea

\noi We assume that $|\bm{g}_2|>|\bm{g}_1|$ and at the same time $t_{||}\neq t_{\rm diag}\neq0$ \& $\Delta\phi+\Delta\eta\neq\pi/2$. The Berry monopoles are defined in $(\Delta\phi+\Delta\eta,\theta_{\rm J},(\Delta\phi_\ell-\pi)\Gamma_\ell-2\epsilon_{{\rm vio},r}\Delta\phi_r)$ space, and are located at positions $(-\pi/2,\theta_c,0)$ with $t_{||}\cos\theta_c+t_{\rm diag}=0$.

\subsection{D. Two Stacked TSC Josephson Junctions}

In the present case, we consider that $\eta_{+,\ell}-\eta_{+,r}=\eta_{-,\ell}-\eta_{-,r}$, as well as $\Delta\eta_{\ell}\neq0,\pi$ and $t_{\perp,\ell}\neq0$. We find:
\bea
\bm{g}_1=\left(0,0,2\sin\left(\frac{\Delta\eta_\ell}{2}\right)t_{\perp,\ell}\cos\Phi_z\right)
\quad{\rm and}\quad
\bm{g}_2=\left(-2\varepsilon(\Delta\phi),2\sin\left(\frac{\Delta\eta_\ell}{2}\right)t_{{\rm diag},\ell},
2\cos\left(\frac{\Delta\eta_\ell}{2}\right)t_{\perp,\ell}\sin\Phi_z,\right)\,.
\eea

\noi The Berry monopoles are in $(\Delta\phi,\Phi_z,t_{{\rm diag},\ell})$ space and are located at $(\Delta\phi_c,0,0)$ and $(\Delta\phi_c,\pi,0)$, with $\varepsilon(\Delta\phi_c)=0$.

\subsection{E. Infinite 1D network of identical TSC Josephson junctions}

Based on Eq.~\eqref{eq:CApump}, we can readily infer the expression for the pumped charge in the case of a 1D ``infinite'' network of coupled identical 1D TSC junctions. As found in Eq.~\eqref{eq:AndreevMode}, for $v_1=-v_2=v$, we have $m=\sqrt{u^2+v^2}q$ and $\tan\theta=u/v$. At the same time, there is only one $\bm{g}$ vector, which implies that for a fixed mode $q$, we find: 
\bea
\Delta Q({\cal A},q)=-e\frac{\Delta\theta}{2\pi}\sum_c{\rm sgn}(v_c)\int_{\cal C}d\varepsilon\ph\delta(\varepsilon-\varepsilon_c)\,.
\eea

\noi We find that the pumping for a full cycle of $\theta$ allows FP pumping. Note that this result does not depend on $q$.

\end{widetext}

\end{document}